\documentclass[%
 onecolumn, secnumarabic,
 amsmath,amssymb,
aps,
]{revtex4}

\usepackage{graphicx}% Include figure files
\usepackage{dcolumn}% Align table columns on decimal point
\usepackage{bm}% bold math
\usepackage{savesym}
\usepackage{txfonts} % this is the font which contain number sign
\usepackage{textcomp}

\providecommand{\U}[1]{\protect\rule{.1in}{.1in}}

\begin{document}

\title{ Quantum deformation of the angular distributions of synchrotron radiation. Emission of particles in the first excited state.}
\author{V. G. Bagrov}
\affiliation{Department of Physics, Tomsk State University, 634050, Tomsk, Russia.}
 \altaffiliation[Also at ]{Tomsk Institute of High Current Electronics, SB RAS, 634034 Tomsk, Russia; Institute of Physics, University of S\~ao Paulo, Brazil}
\email{bagrov@phys.tsu.ru}
\author{A. N. Bourimova}
\altaffiliation[Also at ]{Department of Physics, Tomsk State University, 634050, Tomsk, Russia}
\email{llefrith@yandex.ru}
\author{D. M. Gitman}
\email{gitman@dfn.if.usp.br}
\author{A. D. Levin}
\email{alevin@if.usp.br}
\affiliation{Institute of Physics, University of S\~ao Paulo, Brazil}

\date{\today}% It is always \today, today,

\begin{abstract}
The exact expressions for the characteristics of synchrotron
radiation of charged particles in the first excited state are obtained in
analytical form using quantum theory methods. We performed a detailed analysis of
the angular distribution structure of radiation power and its polarization
for particles with spin 0 and 1/2. It is shown that the exact quantum
calculations lead to results that differ substantially from the predictions
of classical theory.
\end{abstract}

%\pacs{41.60.Ap}% PACS, the Physics and Astronomy
\maketitle

\section{Introduction}

\quad \thinspace \thinspace\ The theory of synchrotron radiation (SR) is now
a very well developed section of theoretical physics and is widely
represented in numerous scientific articles, reviews [1, 2], monographs [3 -
6] and textbooks [7].

In particular, in the classical theory of SR it is possible to find answers to basic questions in clear and analytical form  and to propose algorithms
for numerical simulation of the physical properties of SR: spectral -
angular, spectral, angular distributions and polarization properties.

The quantum theory of SR has also provided a number of significant
achievements which have allowed the scope of classical
theory to be clearly specified. Moreover, with the use
of the quantum approach it has been possible to predict the effects of radiation
induced self-polarization of electron beams and the quantum
excitation of synchrotron oscillations. Both these phenomena
where later confirmed experimentally.

However, up to now we know of very few theoretical results describing the
variation of the angular distributions of SR in regions where the quantum
corrections can no longer be regarded as small. For example, in [8, 9] quantum theory is used to study the SR of non-relativistic particles at low energy levels. It was shown that in the non-relativistic region the influence of quantum corrections is more noticeable the smaller the initial energy level of the particles. Articles [10 - 12] also indicate areas of possible substantial manifestation of the quantum corrections for ultra-relativistic particles. The small number of papers on the theory of SR in regions where quantum corrections become significant is even
more surprising, given that the spectral - angular distribution of SR in
these areas is theoretically described by exact analytical expressions.

The urgency to fill this theoretical gap rises from the fact that the parameters of modern accelerators are very close to the region where quantum corrections should be taken into account. In astrophysics, SR is currently one of the main experimental sources of our knowledge about the physical processes in outer space and, without a doubt, the conditions of space SR can be properly understood only through the use of quantum theory.

Here we investigate the main characteristics of the SR for particles in the first excited state, using exact analytical methods of quantum theory. In
particular, comparative analysis of classical and quantum theory results is performed for angular distributions and radiation polarization.

\section{Radiated Frequencies}

\quad \thinspace \thinspace The  energy $E=m_{0}c^{2}\gamma $  (where, $m_{0}$ is the rest mass, $c$ is the speed of light, and $\gamma $ is the relativistic factor) of a spinless particle (boson) subjected to an external constant and uniform magnetic field with intensity $H>0$ in the absence of motion along the field is given by [3 - 6]
\begin{equation}
\gamma ^{2}=1+(2n+1)B,\ \ B=\frac{H}{H_{0}},\ \ H_{0}=\frac{m_{0}^{2}c^{3}}{%
|e|\hbar }.  \label{11}
\end{equation}
For a particle with spin 1 / 2 (electron) we have
\begin{equation}
\gamma ^{2}=1+2nB,  \label{12}
\end{equation}%
where, $n=0,\,1,\,2,\,3,...$ corresponds to different energy levels, $e$ is the algebraic value of the particle charge, $\hbar $ is Planck's constant. In what follows, only negatively charged ($e<0$) particles will be considered.

We will conduct a comparative analysis of the radiation characteristics for an electron and a boson that have equal energy (the
same relativistic factor $\gamma $) and the same energy level number $n$,
 which according to (\ref{11}), (\ref{12}), involves various intensities $H$ for the electron and the boson.

It is known [1 - 7] that the spectrum of synchrotron radiation is discrete,
and for bosons and electrons with relativistic factor $\gamma $ and
with the energy level number $n$, the possible frequencies of emitted
photons $\omega ^{b},\,\omega ^{e}$ are determined by the expression
\begin{eqnarray}
\frac{\hbar \,\omega ^{e,b}}{m_{0}c^{2}}&=&\frac{\nu }{\overline{n}}\frac{
\gamma \,\beta ^{2}}{1+\sqrt{1-\frac{\nu }{\overline{n}}\beta ^{2}\sin
^{2}\theta }}, \quad \beta ^{2}=1-\frac{1}{\gamma ^{2}}\,,\overline{n}=\left\{
\begin{array}{cc}
n\ \
\mbox{-- for electron} &  \\
n+\frac{1}{2}\ \
\mbox{-- for boson} &
\end{array}%
\right. \,,  \label{13}
\end{eqnarray}%
where, $\nu =1,\,2,\,3,...,n$ is the number of the radiated harmonic, and $\theta $
is the angle between the external magnetic field and the direction of photon
emission. It is not difficult to establish the inequality $\omega
^{e}>\omega ^{b}$. When we consider $\omega ^{e,b}$ as functions of $\theta
$, then other fixed parameters of these functions are maximized at $\theta
=\pi /2$ (photon moves perpendicular to the magnetic field vector and
parallel to the particle orbit plane) and minimized at $\theta=0 $ (photon moves parallel to the magnetic field).

When a particle is in the first excited state $n=1$, only the first harmonic $\nu =1$ may be emitted and from (\ref{11}) -- (\ref{13}) we find for the electron
\begin{equation}
\frac{\hbar \,\omega ^{e}}{m_{0}c^{2}}=\frac{\gamma \,\beta ^{2}}{1+\sqrt{%
1-\beta ^{2}\sin ^{2}\theta }}\,,\ \ \gamma ^{2}=1+2B\,,\ \ \beta ^{2}=\frac{%
2B}{1+2B}\,,  \label{14}
\end{equation}%
and for the boson
\begin{equation}
\frac{\hbar \,\omega ^{b}}{m_{0}c^{2}}=\frac{2}{3}\frac{\gamma \,\beta ^{2}}{%
1+\sqrt{1-\frac{2}{3}\beta ^{2}\sin ^{2}\theta }}\,,\ \ \gamma ^{2}=1+3B\,,\
\ \beta ^{2}=\frac{3B}{1+3B}\,.  \label{15}
\end{equation}

In what follows, a detailed study of the main characteristics of SR for particles in the first excited state $n=\nu =1$ is performed. In this case, all particle radiation properties are determined by its angular distribution and polarization.

Polarization components of the SR will be labeled by the index $s$ and we
choose a standard [1, 5, 6] labeling method: $s=g=\pm 1$, where  $g=1$ is the right and  $g=-1$ is the left circular polarization component; $s=2,\,3$ is used for $\sigma -$ and $\pi -$ components of linear polarization respectively; $s=0$ denotes the total radiated power (equal to the sum of the two linear or the two circular polarization components).

\section{Analytical expressions for the physical characteristics of the
radiated power}

\quad \thinspace \thinspace In quantum theory, general analytical
expressions for the spectral - angular distribution of the SR are known [1,
5, 6] and expressed in terms of Laguerre functions, with complex dependence
on $n,\,\nu ,\,\beta,\,\theta $. In the special case of $n=\nu
=1$ the general formulae are considerably simplified and the relatively simple expressions which can be obtained are given below.

\subsection{Spinless particle (boson)}

\quad \thinspace \thinspace We introduce an intermediate parameter $%
\overline{x}_{0}$ and an intermediate variable $\overline{x}$
\begin{eqnarray}
\overline{x}_{0}&=&\overline{x}_{0}(\beta )=\frac{\sqrt{3}-\sqrt{3-2\beta ^{2}}}{\sqrt{3}+\sqrt{3-2\beta ^{2}}}\,; \quad \overline{x}=\overline{x}(\beta,\,\theta )=\frac{\sqrt{3}-\sqrt{3-2\beta ^{2}\sin ^{2}\theta }}{\sqrt{3}+\sqrt{3-2\beta ^{2}\sin ^{2}\theta }}\,;  \label{21}
\end{eqnarray}%
\begin{equation*}
0\leqslant \overline{x}_{0}(\beta )\leqslant 2-\sqrt{3}\approx
0,26794919\,;\ \ 0\leqslant \overline{x}(\beta ,\,\theta )\leqslant
\overline{x}_{0}(\beta ).
\end{equation*}

With this notation the boson angular distribution of radiated power $W_{s}^{b}$ can be written as
\begin{eqnarray}
\frac{dW_{s}^{b}(\beta ;\theta )}{d\Omega }&=&\frac{Q_{0}A(\beta )}{54}(1+%
\overline{x})^{3}e^{-\overline{x}}\,\varphi _{s}^{b}(\beta ;\theta );\quad
Q_{0}=\frac{e^{2}m_{0}^{2}c^{3}}{\hbar ^{2}},\quad  A(\beta )=\frac{\beta ^{6}%
}{1-\beta ^{2}}=\frac{(\gamma ^{2}-1)^{3}}{\gamma ^{4}}; \nonumber\\
\varphi _{g}^{b}(\beta ;\theta )&=&\frac{1}{2}\varphi _{0}^{b}(\beta ;\theta)+g(1+\overline{x})\cos \theta ,\quad
\varphi _{2}^{b}(\beta ;\theta )=1-\overline{x},\quad \varphi _{3}^{b}(\beta ;\theta )=\frac{(1+\overline{x})^{2}\cos ^{2}\theta }{1-\overline{x}},\nonumber\\
\varphi _{0}^{b}(\beta ;\theta )&=&\varphi _{2}^{b}(\beta ;\theta )+\varphi_{3}^{b}(\beta ;\theta )=\varphi _{-1}^{b}(\beta ;\theta )+\varphi_{1}^{b}(\beta ;\theta ); \quad
\ \ d\Omega=\sin \theta d\theta \,,\ \ g=\pm 1.  \label{22}
\end{eqnarray}

From (\ref{22}), taking into account (\ref{21}) and integrating with respect to $\theta\,\,(0\leqslant \theta \leqslant \pi )$ one obtains the following expression for the total power $W_{0}^{b}(\beta )$ radiated by a boson
\begin{eqnarray}
W_{0}^{b}(\beta )&=&\frac{4\,Q_{0}A(\beta )}{81}f^{b}(\beta ),\quad
f^{b}(\beta)=\frac{3(1+\overline{x}_{0})^{2}}{8}f_{0}^{b}(\overline{x}_{0}),\ \
f_{0}^{b}({x})=f_{2}^{b}(x)+f_{3}^{b}(x).  \label{23}
\end{eqnarray}

From (\ref{22}) and (\ref{21}), it also follows that the power of the
polarization components $W_{s}^{b\,(+)}(\beta )$ radiated in the upper half plane $0\leqslant \theta \leqslant \pi /2$ can be represented as
\begin{eqnarray}
W_{s}^{b\,(+)}(\beta )&=&\frac{1}{2}W_{0}^{b}(\beta )q_{s}^{b}(\beta ),\quad
q_{2}^{b}(\beta )+q_{3}^{b}(\beta )=q_{-g}^{b}(\beta )+q_{g}^{b}(\beta)=1,\,\,(g=\pm 1),  \label{24}
\end{eqnarray}%
where the following notation is used
\begin{equation}
q_{g}^{b}(\beta )=\frac{1}{2}+g\,\frac{f_{1}^{b}(\overline{x}_{0})}{%
f_{0}^{b}(\overline{x}_{0})}\,,\ \ q_{2}^{b}(\beta )=\frac{f_{2}^{b}(%
\overline{x}_{0})}{f_{0}^{b}(\overline{x}_{0})}\,,\ \ q_{0}^{b}(\beta )=1.
\label{25}
\end{equation}

Functions $f_{k}^{b}(x),\,\,(k=1,\,2,\,3)$, introduced in (\ref{23}) -- (\ref{25}), can be represented as integrals of elementary functions, depending on a parameter (the function $f_{1}^{b}(x)$ is elementary itself, however, the functions $f_{2}^{b}(x),\,f_{3}^{b}(x)$ can not be expressed in terms of elementary functions)
\begin{eqnarray}
f_{1}^{b}(x)&=&\int_{0}^{1}(1-x^{2}y^{2})\,\,e^{-xy}\,dy=\frac{(1+x)^{2}e^{-x}-1}{x}\,,\ \ f_{1}^{b}(0)=1,\quad
f_{2}^{b}(x)=\int_{0}^{1}\frac{(1+xy)(1-xy)^{2}}{\sqrt{(1-y)(1-x^{2}y)}}\,\,e^{-xy}\,dy\,,\ \ f_{2}^{b}(0)=2\,\nonumber\\
f_{3}^{b}(x)&=&\int_{0}^{1}(1+xy)\sqrt{(1-y)(1-x^{2}y)}\,\,e^{-xy}\,dy,\quad
f_{3}^{b}(0)=\frac{2}{3},\quad q_{2}^{b}(0)=\frac{3}{4},\ \ q_{g}^{b}(0)=\frac{4+3g}{8}\,.\ \ \ \ \label{26}
\end{eqnarray}

Quantities $q_{s}^{b}(\beta )$ determine the contribution of the $\sigma $--component of polarization to the total radiation in the upper half-plane, which determines the degree of polarization in the upper half-plane.

For numerical calculations of the functions $f_{2}^{b}(x),\,f_{3}^{b}(x)$ it
is convenient to perform the following substitution in the integrals (\ref%
{26})
\begin{eqnarray}
y=\frac{1-t^{2}}{1-x^{2}t^{2}},\ \ 0\leqslant t\leqslant 1;\ \ dy&=&-\frac{2(1-x^{2})tdt}{(1-x^{2}t^{2})^{2}},\quad
\sqrt{(1-y)(1-x^{2}y)}=\frac{(1-x^{2})t}{1-x^{2}t^{2}},  \label{27}
\end{eqnarray}%
after which we have
\begin{eqnarray}
f_{2}^{b}(x)=2(1+x)(1-x)^{2}
\int_{0}^{1}\frac{(1-xt^{2})(1+xt^{2})^{2}}{(1-x^{2}t^{2})^{4}}\exp \left[ -\frac{x(1-t^{2})}{1-x^{2}t^{2}}\right]dt,
\label{28}
\end{eqnarray}%
\begin{eqnarray}
f_{3}^{b}(x)=2(1+x)(1-x^{2})^{2}
\int_{0}^{1}\frac{(1-xt^{2})t^{2}}{(1-x^{2}t^{2})^{4}}\exp \left[ -\frac{x(1-t^{2})}{1-x^{2}t^{2}}\right] dt.
\label{29}
\end{eqnarray}%
When $|x|<1$ the integrand in (\ref{28}) and (\ref{29}) remains finite
for all values of the integration variable, whereas in (\ref{26}) the
second integral is improper.

For the power of polarization components $W_{s}^{b\,(-)}(\beta
)\,\,(s=-1,\,1,\,2,\,3)$ emitted in the lower half plane $\pi /2\leqslant \theta
\leqslant \pi $, we obtain
\begin{eqnarray}
W_{2}^{b\,(-)}(\beta )&=&W_{2}^{b\,(+)}(\beta )\,,\ \ W_{3}^{b\,(-)}(\beta)=W_{3}^{b\,(+)}(\beta ), \quad  W_{g}^{b\,(-)}(\beta )=W_{-g}^{b\,(+)}(\beta).  \label{210}
\end{eqnarray}

Using expression (\ref{23}), the angular distribution of radiated power (%
\ref{22}) can be represented as
\begin{eqnarray}
\frac{dW_{s}^{b}(\beta ;\theta )}{d\Omega }&=&W_{0}^{b}(\beta )p_{s}^{b}(\beta
;\theta ), \quad
p_{s}^{b}(\beta ;\theta )=\frac{(1+\overline{x})^{3}e^{-\overline{x}}\,\varphi _{s}^{b}(\beta ;\theta )}{(1+\overline{x}%
_{0})^{2}f_{0}^{b}(\overline{x}_{0})}\,;  \label{211}
\end{eqnarray}%
Functions $p_{s}^{b}(\beta ;\theta )$ have the properties
\begin{eqnarray}
p_{s}^{b}(\beta ;\theta )=p_{s}^{b}(\beta ;\pi -\theta ),\, (s=0,\,2,\,3);\
\ p_{g}^{b}(\beta ;\theta )=p_{-g}^{b}(\beta ;\pi -\theta ).  \notag
\end{eqnarray}%
Here the quantities $p_{s}^{b}(\beta ;\theta )\,d\Omega $ determine the
contribution to the total radiated power of the $s$-- polarization component emitted within a small solid angle $d\Omega $ around the direction defined by angle $\theta $, which means that $p_{s}^{b}(\beta ;\theta )$ are distribution functions of $\theta $. We give some special values of the functions $p_{s}^{b}(\beta ;\theta )$
\begin{eqnarray}
p_{0}^{b}(\beta ;0)=p_{1}^{b}(\beta ;0)=2p_{2}^{b}(\beta
;0)=2p_{3}^{b}(\beta ;0)=\frac{2}{(1+\overline{x}_{0})^{2}f_{0}^{b}(%
\overline{x}_{0})}, \nonumber\\
p_{-1}^{b}(\beta ;0)=0;\, p_{0}^{b}\left( \beta ;\frac{\pi }{2}\right)=p_{2}^{b}\left( \beta ;\frac{%
\pi }{2}\right) =2p_{g}^{b}\left( \beta ;\frac{\pi }{2}\right) =\frac{1-%
\overline{x}_{0}^{\,2}}{e^{\,\overline{x}_{0}}f_{0}^{b}(\overline{x}_{0})},\quad
p_{3}^{b}\left( \beta ;\frac{\pi }{2}\right)=0.  \label{212}
\end{eqnarray}

The following relation is obvious
\begin{eqnarray}
q_{s}^{b}(\beta )=2\int_{0}^{\frac{\pi }{2}}p_{s}^{b}(\beta ;\theta
)\,d\Omega \,.  \label{213}
\end{eqnarray}

Let's introduce the functions
\begin{eqnarray}
q_{s}^{b}(\beta ;\theta )=\frac{p_{s}^{b}(\beta ;\theta )}{p_{0}^{b}(\beta
;\theta )}=\frac{\varphi _{s}^{b}(\beta ;\theta )}{\varphi _{0}^{b}(\beta
;\theta )}.  \label{214}
\end{eqnarray}%
These functions have the properties
\begin{eqnarray}
q_{2}^{b}(\beta ;\theta )+q_{3}^{b}(\beta ;\theta )=q_{g}^{b}(\beta ;\theta
)+q_{-g}^{b}(\beta ;\theta )=q_{0}^{b}(\beta ;\theta )=1;  \notag
\end{eqnarray}%
\begin{eqnarray}
q_{s}^{b}(\beta ;\theta )=q_{s}^{b}(\beta ;\pi -\theta )\ \ (s=2,\,3);\ \
q_{g}^{b}(\beta ;\theta )=q_{-g}^{b}(\beta ;\pi -\theta ),  \notag
\end{eqnarray}%
from which it follows that it is sufficient to study the behavior of $%
q_{1}^{b}(\beta ;\theta )$ and $q_{2}^{b}(\beta ;\theta )$ only. The functions $%
q_{s}^{b}(\beta ;\theta )$ determine the contribution of the $s$--polarization component to the angular distribution of radiation in the direction given by $\theta$, and determine the degree of polarization for each fixed $\beta $ and $\theta $. We find some particular
values of the functions $q_{s}^{b}(\beta ;\theta )$
\begin{eqnarray}
q_{2}^{b}(\beta ;0)&=&\frac{1}{2}\,,\ \ q_{g}^{b}(\beta ;0)=\frac{1+g}{2},
\quad
q_{2}^{b}\left( \beta ;\frac{\pi }{2}\right)=1\,,\ \ q_{g}^{b}\left(
\beta ;\frac{\pi }{2}\right) =\frac{1}{2}\,.  \label{215}
\end{eqnarray}

Note that for functions $f^{b}(\beta ),\,q_{s}^{b}(\beta ;\theta
),\,p_{s}^{b}(\beta ;\theta ),\,q_{s}^{b}(\beta ;\theta )$ point $\beta =1$
is not singular, and at this point there exist continuous derivatives with respect to $\beta
$ for all these functions.

\subsection{Spinor particle (electron)}

We introduce an intermediate parameter $x_{0}$ and an intermediate variable $%
x$
\begin{eqnarray}
x_{0}=x_{0}(\beta )&=&\frac{1-\sqrt{1-\beta ^{2}}}{1+\sqrt{1-\beta ^{2}}}=\frac{\gamma -1}{\gamma +1};\quad
x=x(\beta ,\,\theta )=\frac{1-\sqrt{1-\beta ^{2}\sin ^{2}\theta }}{1+\sqrt{1-\beta ^{2}\sin ^{2}\theta }}\,;
\label{216}
\end{eqnarray}%
\begin{eqnarray*}
0\leqslant x_{0}(\beta )\leqslant 1\,;\ \ 0\leqslant x(\beta ,\,\theta
)\leqslant x_{0}(\beta )\,.
\end{eqnarray*}%
For an electron, the SR characteristics depend also on spin
orientation. We will consider the transverse [1, 3-7] orientation of electron spin, defining it with the spin quantum number $\zeta =\pm 1$. Let $\zeta =1$ correspond to the orientation of electron spin in
the initial state along the direction of the magnetic field and $\zeta =-1$
correspond to the orientation of electron spin in the initial state
opposite to the direction of the magnetic field. It is known [1,\thinspace
5,\thinspace 6] that, in the case under consideration, in the final state (for which $n^{\prime }=0$) the electron spin can be aligned only against the direction of the magnetic field, $\zeta ^{\prime }=-1$. Thus, for $\zeta =-1$
the transition to the ground state occurs without spin flip, and for $\zeta =1$ the transition is necessarily accompanied by spin flip. With this in mind, the electron angular distribution of radiated power $W_{s}^{e}$ can be
written as
\begin{eqnarray}
\frac{dW_{s}^{e}(\zeta ;\beta ;\theta )}{d\Omega }&=&d(\zeta ;\,\beta )\frac{%
Q_{0}A(\beta )}{16(1+x_{0})}\,\frac{(1+x)^{3}e^{-x}}{1-x}\,\varphi
_{s}^{e}(\zeta ;\beta ;\theta );  \nonumber\\
\varphi _{2}^{e}(-1;\beta ;\theta )&=&\varphi _{3}^{e}(1;\beta ;\theta)=1-x_{0}x ,\quad
\varphi _{2}^{e}(1;\beta ;\theta )=\varphi _{3}^{e}(-1;\beta ;\theta )=\frac{(1+x)^{2}\cos ^{2}\theta }{1-x_{0}x},\nonumber\\
\varphi _{0}^{e}(\zeta ;\beta ;\theta )&=&\varphi _{0}^{e}(\beta ;\theta)=\varphi _{2}^{e}(-1;\beta ;\theta )+\varphi _{3}^{e}(-1;\beta ;\theta)
=\varphi _{2}^{e}(1;\beta ;\theta )+\varphi _{3}^{e}(1;\beta ;\theta ),\nonumber\\
\varphi _{g}^{e}(\zeta ;\beta ;\theta )&=&\varphi _{g}^{e}(\beta ;\theta )=\frac{\varphi _{0}^{e}(\beta ;\theta )}{2}+g(1+x)\cos \theta ,\quad
g=\pm 1. \label{217}
\end{eqnarray}%
Here, we introduce the function
\begin{eqnarray}
d(\zeta ;\,\beta )=\frac{1-\zeta +x_{0}(1+\zeta )}{2}=\left\{
\begin{array}{cc}
x_{0}\ \ \mbox{when}\ \ \zeta =1; &  \\
\,\,\,\,\,1\ \ \mbox{when}\ \ \zeta =-1\,. &
\end{array}%
\right.  \label{218}
\end{eqnarray}

By integrating expressions (\ref{217}) with respect to $\theta\,\,(0\leqslant \theta \leqslant \pi )$ and taking (\ref{216}) into account we obtain the following expression for total radiated power of an electron
\begin{eqnarray}
W_{0}^{e}(\zeta ;\,\beta )&=&d(\zeta ;\,\beta )\frac{Q_{0}A(\beta )}{6}%
f^{e}(\beta ),\quad
f^{e}(\beta )=\frac{3(1+x_{0})}{8}f_{0}^{e}(x_{0}),\ \
f_{0}^{e}({x})=f_{2}^{e}(x)+f_{3}^{e}(x).  \label{219}
\end{eqnarray}%
The power of the polarization
components $W_{s}^{e\,(+)}(\zeta ;\,\beta )$ emitted by an electron in the upper half plane $0\leqslant \theta \leqslant \pi /2$ can be represented similarly to (\ref{24}), (\ref{25}),
\begin{eqnarray}
W_{s}^{e\,(+)}(\zeta ;\,\beta )&=&\frac{1}{2}W_{0}^{e}(\zeta ;\,\beta)q_{s}^{e}(\zeta ;\,\beta ),\quad
q_{g}^{e}(\zeta ;\,\beta )\equiv q_{g}^{e}(\beta ),q_{2}^{e}(\zeta ;\,\beta )=q_{3}^{e}(-\zeta ;\,\beta ),\nonumber\\
q_{2}^{e}(\zeta ;\,\beta )&+&q_{3}^{e}(\zeta ;\,\beta )=q_{-g}^{e}(\beta )+q_{g}^{e}(\beta)=q_{0}^{e}(\zeta ;\,\beta )=1,  \label{220}
\end{eqnarray}
where we have the following relations
\begin{eqnarray}
q_{1}^{e}(\beta )=\frac{1}{2}+\frac{f_{1}^{e}(x_{0})}{f_{0}^{e}(x_{0})}\,,\
\ q_{2}^{e}(\zeta ;\,\beta )=\frac{1+\zeta -2\zeta q_{2}^{e}(\beta )}{2},\quad
q_{2}^{e}(\beta )=\frac{f_{2}^{e}(x_{0})}{f_{0}^{e}(x_{0})}.  \label{221}
\end{eqnarray}%
The functions $f_{2}^{e}(x)\,\,f_{3}^{e}(x)$, introduced in (\ref{218}) -- (%
\ref{221}), are expressed as integrals depending on a parameter (but can not
be expressed through elementary functions), but the function $f_{1}^{e}(x)$
is elementary
\begin{eqnarray}
f_{1}^{e}(x)&=&\frac{2-(2+x)\exp (-x)}{x}\,;\ \ 0\leqslant x\leqslant 1;\nonumber\\
f_{2}^{e}(x)&=&\int_{0}^{1}(1+xy)\exp (-xy)\sqrt{\frac{1-x^{2}y}{1-y}}dy=
2(1+x)(1-x^{2})\int_{0}^{1}\frac{1-xt^{2}}{(1-x^{2}t^{2})^{3}}\exp \left[ -%
\frac{x(1-t^{2})}{1-x^{2}t^{2}}\right] dt; \nonumber\\
f_{3}^{e}(x)&=&\int_{0}^{1}(1+xy)\exp (-xy)\sqrt{\frac{1-y}{1-x^{2}y}}dy=2(1+x)(1-x^{2})
\int_{0}^{1}\frac{(1-xt^{2})t^{2}}{(1-x^{2}t^{2})^{3}}\exp\left[-\frac{x(1-t^{2})}{1-x^{2}t^{2}}\right] dt.\ \ \ \label{222}
\end{eqnarray}%
In the vicinity of the boundary points $x=0;\,1$ we find

\begin{eqnarray}
f_{1}^{e}(x)&\approx& 1-\frac{x^{2}}{6},\quad f_{2}^{e}(x)\approx 2\left( 1-\frac{13}{15}x^{2}\right) ,\quad
f_{3}^{e}(x)\approx\frac{2}{3}\left( 1+\frac{11}{35}x^{2}\right) ;\ \ x\ll 1. \quad
f_{1}^{e}(x)\approx 2-\frac{3}{e}+\left( 2-\frac{5}{e}\right) (1-x),\nonumber\\
f_{2}^{e}(x)&\approx& 2-\frac{3}{e}-\frac{4}{e}(1-x)\ln (1-x),\quad
f_{3}^{e}(x)\approx 2-\frac{3}{e}+\frac{2}{e}(1-x)\ln (1-x);\quad 0<1-x\ll 1,\nonumber\\
q_{2}^{e}(0)&=&\frac{3}{4},\quad q_{g}^{e}(0)=\frac{4+3g}{8};
q_{2}^{e}(1)=\frac{1}{2},\quad q_{g}^{e}(1)=\frac{1+g}{2}.  \label{223}
\end{eqnarray}

For the power of the polarization components $W_{s}^{e\,(-)}(\zeta ;\,\beta )$ emitted in the lower half plane $\pi /2\leqslant \theta \leqslant \pi $, the relations (\ref{210}) are valid.

By analogy with (\ref{211}), for an electron one can also introduce $p_{s}^{e}(\zeta ;\,\beta ;\theta )$ -- the distribution of radiated power as a function of the angle $\theta $. Then expression (\ref{217}) for radiated power $W_{s}^{e}$ for an electron takes the following form
\begin{eqnarray}
\frac{dW_{s}^{e}(\zeta ;\,\beta ;\theta )}{d\Omega }&=&W_{0}^{e}(\zeta;\,\beta )p_{s}^{e}(\zeta ;\,\beta ;\theta ),\quad
p_{s}^{e}(\zeta ;\,\beta;\theta )=\frac{(1+x)^{3}e^{-x}\,\varphi _{s}^{e}(\zeta ;\,\beta ;\theta )}{(1-x)(1+x_{0})^{2}f_{0}^{e}(x_{0})};\quad
p_{2}^{e}(\zeta ;\beta ;\theta )=p_{3}^{e}(-\zeta ;\beta ;\theta ),\nonumber\\
p_{0}^{e}(\zeta ;\beta ;\theta )&\equiv& p_{0}^{e}(\beta ;\theta ),\quad p_{g}^{e}(\zeta ;\beta ;\theta )\equiv p_{g}^{e}(\beta ;\theta );\quad
p_{2}^{e}(-1;\beta ;\theta )=p_{3}^{e}(1;\beta ;\theta )=p_{2}^{e}(\beta;\theta ),\quad
p_{3}^{e}(-1;\beta ;\theta )=p_{2}^{e}(1;\beta ;\theta)=p_{3}^{e}(\beta ;\theta );\nonumber\\
p_{s}^{e}(\zeta ;\beta ;\theta )&=&p_{s}^{e}(\zeta ;\beta ;\pi -\theta ),\quad(s=0,\,2,\,3),\quad
p_{g}^{e}(\beta ;\theta )=p_{-g}^{e}(\beta ;\pi -\theta ).\ \ \label{224}
\end{eqnarray}%
We present some particular values of the functions $%
p_{s}^{e}(\zeta ;\beta ;\theta )$
\begin{eqnarray}
p_{0}^{e}(\beta ;0)= p_{1}^{e}(\beta ;0)=2p_{2}^{e}(\zeta ;\beta ;0)=\frac{2}{(1+x_{0})^{2}f_{0}^{e}(x_{0})},\quad p_{-1}^{e}(\beta ;0)= 0;\nonumber\\
p_{0}^{e}\left( \beta ;\frac{\pi }{2}\right)=p_{2}^{e}\left( \beta ;\frac{\pi }{2}\right) =2p_{g}^{e}\left( \beta ;\frac{\pi }{2}\right)=\frac{(1+x_{0})^{2}}{e^{\,x_{0}}f_{0}^{e}(x_{0})}, \quad
p_{3}^{e}\left( \beta ;\frac{\pi }{2}\right) =0.\ \ \label{225}
\end{eqnarray}%
For an electron, one can also introduce (by analogy with (\ref{214})) the functions
{$q_{s}^{e}(\zeta ;\beta ;\theta )$}, which determine the contribution of the $s$--polarization component to the angular distribution of radiation in the direction given by $\theta $, and a degree of radiation polarization for each fixed\ $\zeta, \beta, \theta $. Taking into account (\ref{217}) and (\ref{224}) we find
%\begin{eqnarray}
\begin{eqnarray}
q_{s}^{e}(\zeta ;\beta ;\theta )=\frac{p_{s}^{e}(\zeta ;\,\beta ;\,\theta )}{p_{0}^{e}(\zeta ;\,\beta ;\,\theta )}&=&\frac{\varphi _{s}^{e}(\zeta ;\,\beta;\,\theta )}{\varphi _{0}^{e}(\beta ;\,\theta )};\quad q_{g}^{e}(\zeta;\beta ;\theta )\equiv q_{g}^{e}(\beta ;\theta ), \quad
q_{2}^{e}(-1;\beta ;\theta )=q_{2}^{e}(\beta ;\theta ),\nonumber \\[0.7\baselineskip]
q_{2}^{e}(\zeta
;\beta ;\theta )=q_{3}^{e}(-\zeta ;\beta ;\theta ),&&
q_{2}^{e}(\zeta ;\beta ;\theta )+q_{3}^{e}(\zeta ;\beta ;\theta
)=q_{g}^{e}(\beta ;\theta )+q_{-g}^{e}(\beta ;\theta )=q_{0}^{e}(\zeta
;\beta ;\theta )=1,\nonumber\\[0.7\baselineskip]
q_{s}^{e}(\zeta ;\beta ;\theta )&=&q_{s}^{e}(\zeta ;\beta ;\pi -\theta ),\, (s=2,\,3),\quad
q_{g}^{e}(\beta ;\theta )=q_{-g}^{e}(\beta ;\pi -\theta ). \label{226}
\end{eqnarray}
Below are some particular values of the functions $q_{s}^{e}(\zeta ;\beta ;\theta )$
\begin{eqnarray}
q_{2}^{e}(\beta ;0)&=&\frac{1}{2}\,,\ \ q_{g}^{e}(\beta ;0)=\frac{1+g}{2}\,;\quad
q_{2}^{e}\left( \beta ;\frac{\pi }{2}\right) =1\,,\ \ q_{g}^{e}\left(\beta ;\frac{\pi }{2}\right) =\frac{1}{2}\,.  \label{227}
\end{eqnarray}

\section{The main features of the physical characteristics of the radiated
power}

We will analyze the main features of the various physical characteristics
of the SR for our case, based on the above theoretical expressions.

\textbf{1}. It is important to note that the angular distribution functions $p_{s}^{b}(\beta ;\theta )$ and $p_{s}^{e}(\zeta ;\,\beta
;\theta )$ are finite for all values of $\beta $ and $\theta $ (including
the point $\beta =1$, which is not the case in classical theory). For a
boson it clearly follows from (\ref{211}) and for an electron from (\ref{224}) that for all the functions $p_{s}^{e}(\beta ;\theta )$ there is a finite limit $p_{s}^{e}(\beta \rightarrow 1;\theta )\mid _{\theta \neq \frac{\pi }{2}}=\overline{p}\,_{s}^{e}(\theta )$. The functions $\overline{p}\,_{s}^{e}(\theta )$ have the form
\begin{eqnarray}
\overline{p}\,_{2}^{e}(\theta )=\overline{p}\,_{3}^{e}(\theta )=\frac{1}{2}%
\overline{p}\,_{0}^{e}(\theta )=\frac{\Theta (\theta )}{2e-3},\quad
\overline{p}\,_{g}^{e}(\theta )=\frac{\Theta (\theta )}{2e-3}\left( 1+g\frac{\cos\theta }{|\cos \theta |}\right) ,  \label{31}
\end{eqnarray}

where we use the notation
\begin{eqnarray}
\Theta (\theta )=\frac{1}{(1+|\cos \theta |)^{3}}\exp \left( \frac{2|\cos
\theta |}{1+|\cos \theta |}\right) .  \label{32}
\end{eqnarray}

The function $\Theta (\theta )$ is continuous at any $\theta $ and its
derivative has a finite gap at $\theta =\frac{\pi }{2}$ . In particular, we have
\begin{eqnarray}
\overline{p}\,_{2}^{e}\left( \frac{\pi }{2}\right) =\overline{p}%
\,_{3}^{e}\left( \frac{\pi }{2}\right) =\frac{1}{2}\overline{p}%
\,_{0}^{e}\left( \frac{\pi }{2}\right) =\frac{1}{2e-3},  \label{33}
\end{eqnarray}
whereas from (\ref{223}), (\ref{225}) it follows that
\begin{eqnarray}
p_{2}^{e}\left( \beta \rightarrow 1;\frac{\pi }{2}\right) &=&p_{0}^{e}\left(
\beta \rightarrow 1;\frac{\pi }{2}\right) =\frac{2}{2e-3},\quad
p_{3}^{e}\left( \beta \rightarrow 1;\frac{\pi }{2}\right) =0,  \label{34}
\end{eqnarray}
which indicates the ambiguity of the double limit $\beta \rightarrow 1,\ \
\theta \rightarrow \pi /2$.

\textbf{2}. First we should note the following remarkable feature. As follows from the second and third rows of the formulae (\ref{217}) and (\ref{221}),
the function of angular distribution of the $\sigma $--component of polarization of SR for an electron with spin $\zeta =-1$ coincides exactly with the function of the angular distribution of the $\pi $--linear polarization component of SR for an electron with spin $\zeta =1$.
Accordingly, the function of the angular distribution of the $\pi $--component of linear polarization for SR of an electron with spin $\zeta =-1$ coincides exactly with the function of the angular distribution of the $\sigma $--linear polarization component for SR of an electron with spin $\zeta =1$.

We can say that the angular distributions of linear polarization for an electron with spin $\zeta =-1$ and an electron
with spin $\zeta =1$ "switch places". It is possible that this feature of
the angular distributions for electron SR takes place not only for the initial state $n=1$.

This angular distribution feature of the linear
polarization component of electron SR provides a physical explanation for the following known fact. In classical SR theory and quantum SR theory for
spinless particles, the emission of the $\pi $\--linear polarization component is absent in the orbit plane $(\theta =\frac{\pi }{2})$ (according to these theories the radiation in this direction is completely
linearly polarized and only the $\sigma $--component of linear polarization is emitted). This is not the case for electrons and quantum SR theory
tell us that for spinor particles the emission of the $\pi $--linear polarization component is not equal to zero at $\theta =\frac{\pi }{2}$.

This result was first obtained (but has not been specifically emphasized)
in [13, 14], and in [15] this discrepancy with classical
theory was first noted, but no physical analysis of this fact was carried
out. Only in [16] was such an analysis conducted. In particular, it was
found that non-zero emission of the {$\pi $}--component of linear polarization in the orbit plane is due solely to an electron transition with transverse spin. This fact can be used as a possible indicator of
spin orientation. However, in [16] the possibility of the angular distributions between the linear polarization components "switch
places" depending on the orientation of electron spin was not found.

We established here that the linear polarization components of SR "switch places", depending on the orientation of the transverse electron spin. This fact reveals a physical cause for the presence of the {$\pi $}--linear polarization component of electron SR in the orbit plane.

\textbf{3}. The presence of a factor $d(\zeta ;\,\beta )$, defined by
(\ref{218}), in the expressions for the power of electron SR, (\ref{217}) and (\ref{219}), indicates that an electron with spin $\zeta =1$ (in this
case, the radiation is accompanied by a spin flip) always radiates
$x_{0}(\beta )$ times less than an electron with spin $\zeta =-1$
(radiation without a spin flip). Thus, spin-flip transitions
are always $x_{0}(\beta )$ times less probable than transitions without
spin flip. Since
\begin{eqnarray}
x_{0}(\beta )&\approx& \frac{1}{4}\beta ^{2}\ll 1,\ \ \mbox{when}\,\,\beta \ll 1;\quad
x_{0}(\beta )\approx 1,\ \ \mbox{when}\,\,1-\beta ^{2}\ll 1,  \label{35}
\end{eqnarray}%
then in the non relativistic approximation only electrons with spin $\zeta
=-1 $ emit, whereas electrons with spin $\zeta =1$ in practice do not emit
(remain in a quasi-stable state). In the relativistic case $\gamma \gg 1,$
the dependence of SR power on the spin orientation disappears; transitions
with and without spin flip are equiprobable (however exchange of places
between $\sigma $-- and $\pi $--components of SR linear polarization, which
depends on the initial spin state, is preserved at all energies).

\textbf{4}. Using (\ref{23}) and (\ref{219}) we find the ratio of
the total radiated power of electron SR to the total radiated power of
boson SR, when an electron and a boson in the initial state are at energy
level $n=1$ and have the same energy
\begin{eqnarray}
k(\zeta ;\,\beta )&=&\frac{W_{0}^{e}(\zeta ;\,\beta )}{W_{0}^{b}(\beta )};\ \
k(-1;\,\beta )=\frac{27}{8}\,\frac{f^{e}(\beta )}{f^{b}(\beta )}, \quad
k(1;\,\beta )=x_{0}k(-1;\,\beta )\leqslant k(-1;\,\beta ).\ \  \label{36}
\end{eqnarray}

Numerical calculation, for which the results are summarized in Table 1,
shows that $3,375\leqslant k(-1;\,\beta )<3,717$. For the function $%
k(1;\,\beta )$ we have: $k(1;\,\beta )<1$ when $\beta <\beta _{0}\approx
0,8199913\ \ (\gamma <\gamma _{0}\approx 1,7471034)$ and $k(1;\,\beta )>1$
when $\beta >\beta _{0}\ \ (\gamma >\gamma _{0})$.

Thus, an electron with spin antiparallel to the field emits at all
energies almost four times more than a boson, and an electron with spin
along the field begins to emit more than a boson only in the relativistic
domain. The radiated power for electrons significantly depends on spin
in the weakly relativistic region, but with increasing particle energy, this
dependence disappears.
\begin{figure*}[!]
\begin{minipage}[!]{8.1cm}
\centering
{Table 1. Calculated values}
\\
\centering
\begin{tabular}{c|c|c|c|c}
\hline
\hline
$\beta$ & $f^{b}(\beta )$ &  $f^{e}(\beta )$ & $k(-1;\,\beta )$ & $k(1;\,\beta )$\\
\hline
\hline
0.0 & 1.00000 & 1.00000 & 3.37500 & 0.00000\\
0.1 & 1.00167 & 1.00002 & 3.37783 & 0.00849\\
0.2 & 1.00673 & 1.01016 & 3.38651 & 0.03456\\
0.3 & 1.01532 & 1.02333 & 3.40164 & 0.08019\\
0.4 & 1.02769 & 1.04272 & 3.42438 & 0.14917\\
0.5 & 1.04423 & 1.06948 & 3.45661 & 0.24817\\
0.6 & 1.06550 & 1.10543 & 3.50147 & 0.38905\\
0.7 & 1.09533 & 1.15920 & 3.56422 & 0.59438\\
0.8 & 1.12581 & 1.21899 & 3.65435 & 0.91359\\
0.9 & 1.16774 & 1.31125 & 3.78977 & 1.48887\\
1.0 & 1.22085 & 1.34454 & 3.71695 & 3.71695\\
\hline
\hline
\end{tabular}
\end{minipage}
\hfill
\begin{minipage}[!]{8.1cm}
\centering
\includegraphics[width=6cm]{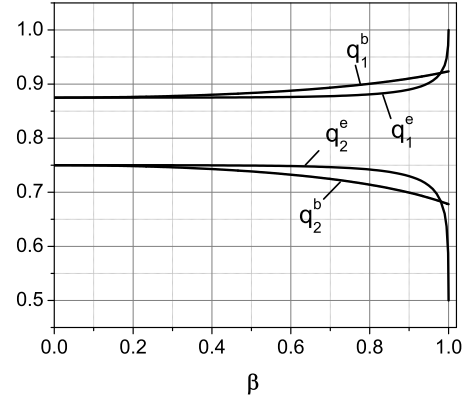}
  \caption{Functions $q_{s}^{b}(\beta), q_{s}^{e}(\beta), (s=1, 2).$}
\end{minipage}
\end{figure*}

\textbf{5}. Consider the energy dependence of the functions $q_{s}^{b}(\beta ),\,\,q_{s}^{e}(\beta )\,\,\,(s=1,\,2)$, which determine the
polarization of the total radiation. Graphs of these functions are shown in the Figure 1.

A quite unusual feature of the case under consideration is that for the
ultra-relativistic electron $(\beta \rightarrow 1)$, according to the precise limits (\ref{223}), just the right-handed polarized radiation remains in the upper half plane (in the ultra-relativistic limit, the left-handed polarized radiation in the upper half plane is vanishingly small compared with the right-handed polarized radiation). In contrast, in the lower half plane the right-handed polarized radiation is vanishingly small, and only the left-handed polarization is emitted. The preferential linear polarization is completely absent. For bosons (also in classical theory), there is no such phenomenon.

In the entire field of finite energy, the degree of right-handed circular
polarization in the upper half plane increases with energy for both the
electron and for the boson (whereas in classical theory it decreases).
The degree of linear polarization of the electron and boson
decreases with increasing energy (in classical theory it increases).

\textbf{6}. The structure of the angular distribution functions $p_{s}^{b}(\beta ;\theta )$ and $p_{s}^{e}(\beta ;\theta )$ is illustrated in
Figures 2 -- 9, which demonstrate the evolution of these functions with particle energy.

\begin{figure*}[!]
\begin{minipage}[!]{8.1cm}
\centering
\includegraphics[width=8cm]{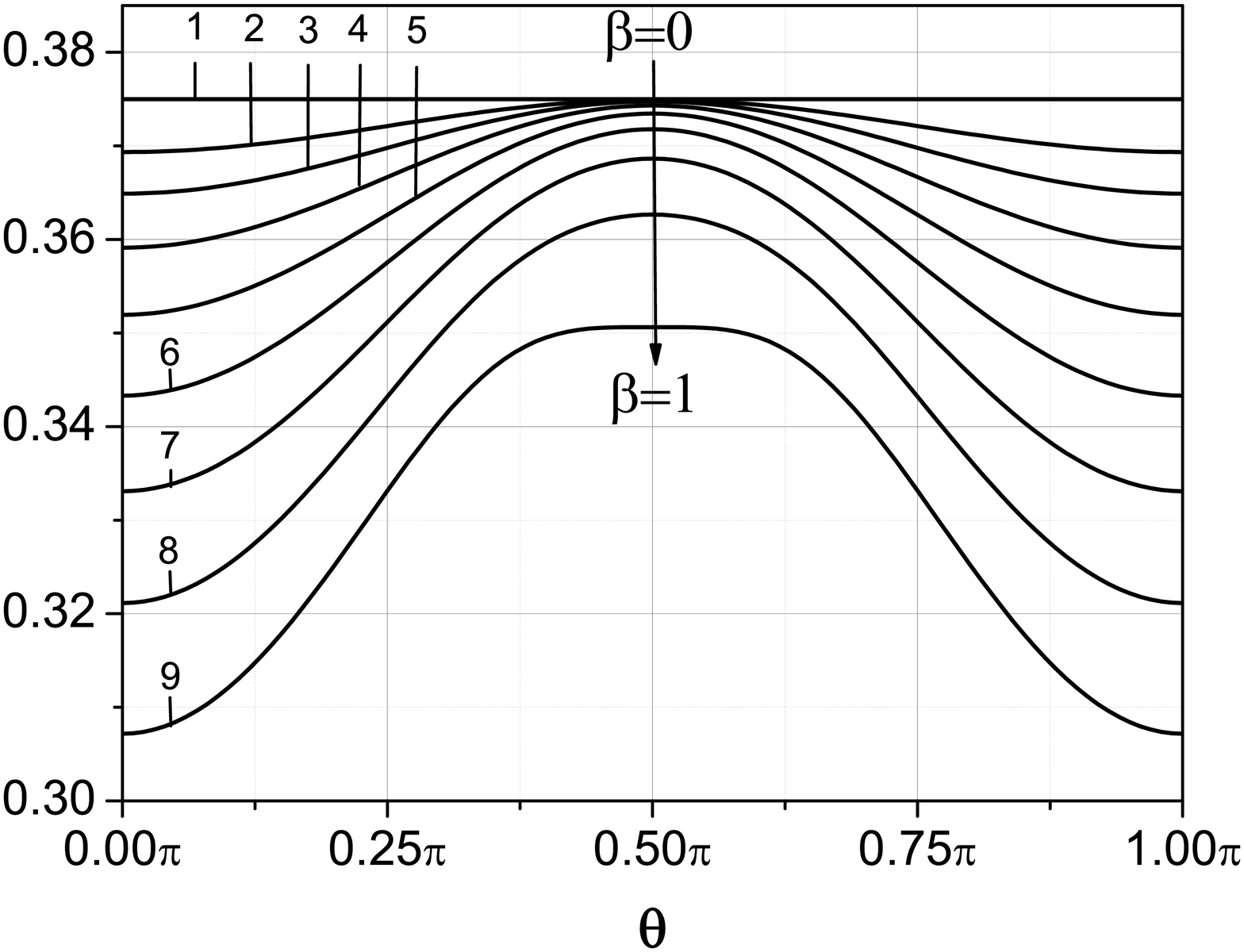}
\caption{Functions $p_{2}^{b}(\beta ;\theta )$. Curve numbers correspond to increasing values
of $\beta$ in the following order:\\ $\{0.0, 0.3, 0.4, 0.5, 0.6, 0.7, 0.8, 0.9, 1.0\}$.}
\end{minipage}
\hfill
\begin{minipage}[!]{8.1cm}
\centering
\includegraphics[width=8cm]{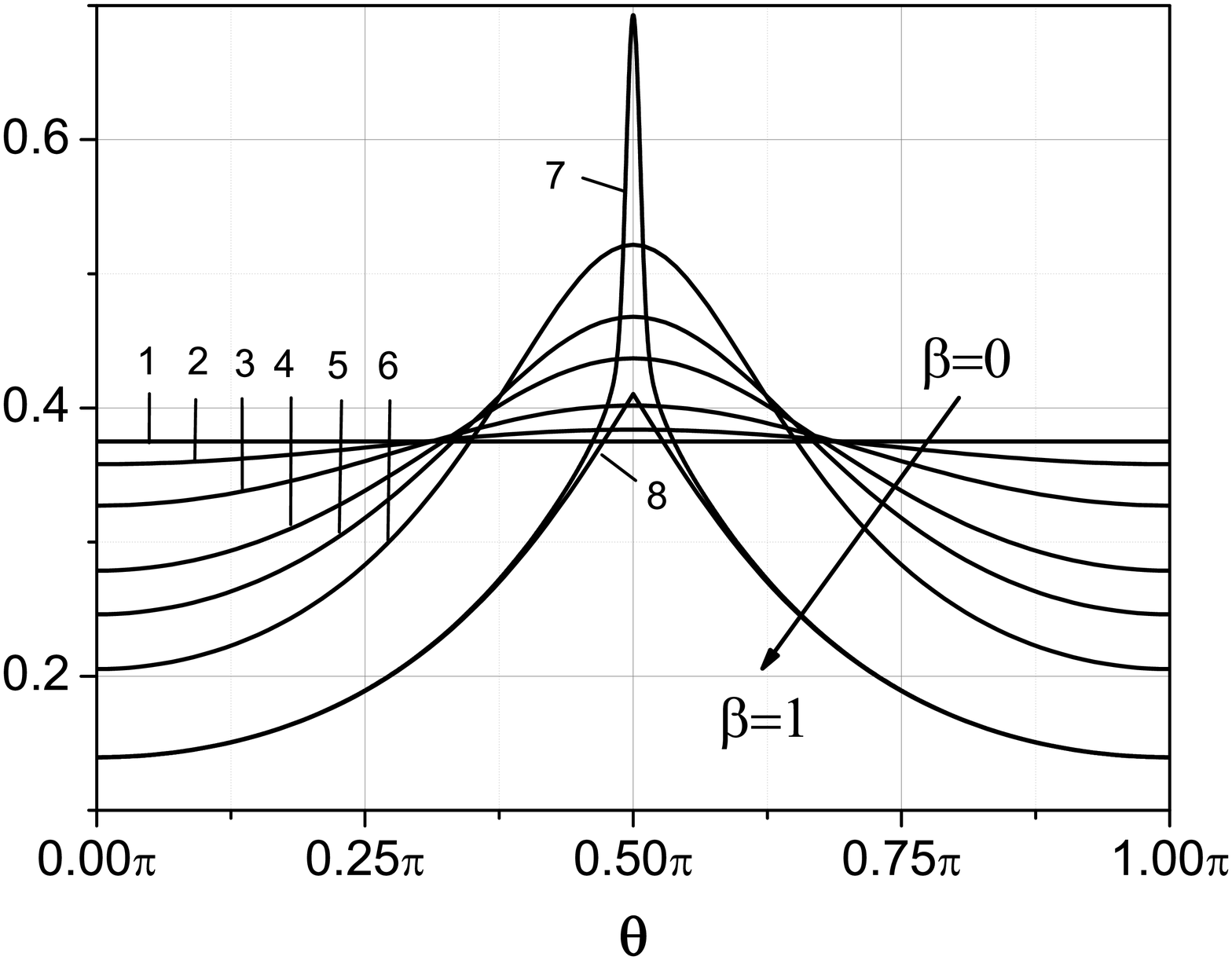}
\caption{Functions $p_{2}^{e}(\beta ;\theta )$. Curve numbers correspond to increasing values
of $\beta$ in the following order:\\ $\{ 0.0, 0.3, 0.5, 0.6, 0.8, 0.9, 0.99999, 1\}$.}
\end{minipage}
\end{figure*}

\begin{figure*}[!]
\begin{minipage}[!]{8.1cm}
\centering
\includegraphics[width=8cm]{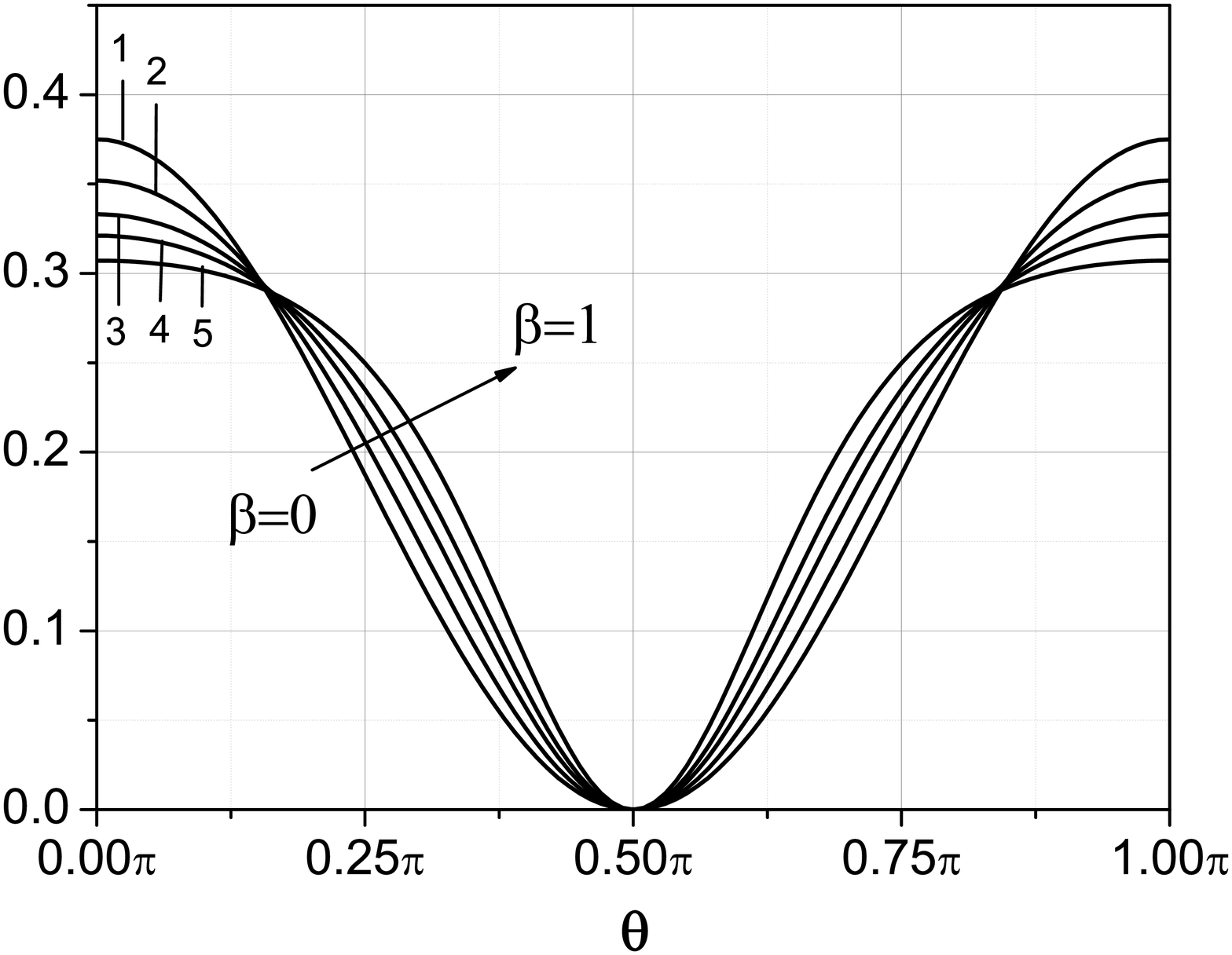}
\caption{Functions $p_{3}^{b}(\beta ;\theta )$. Curve numbers correspond to increasing values
of $\beta$ in the following order:\\ $\{0.0, 0.6, 0.8, 0.9, 1.0\}$.}
\end{minipage}
\hfill
\begin{minipage}[!]{8.1cm}
\centering
\includegraphics[width=8cm]{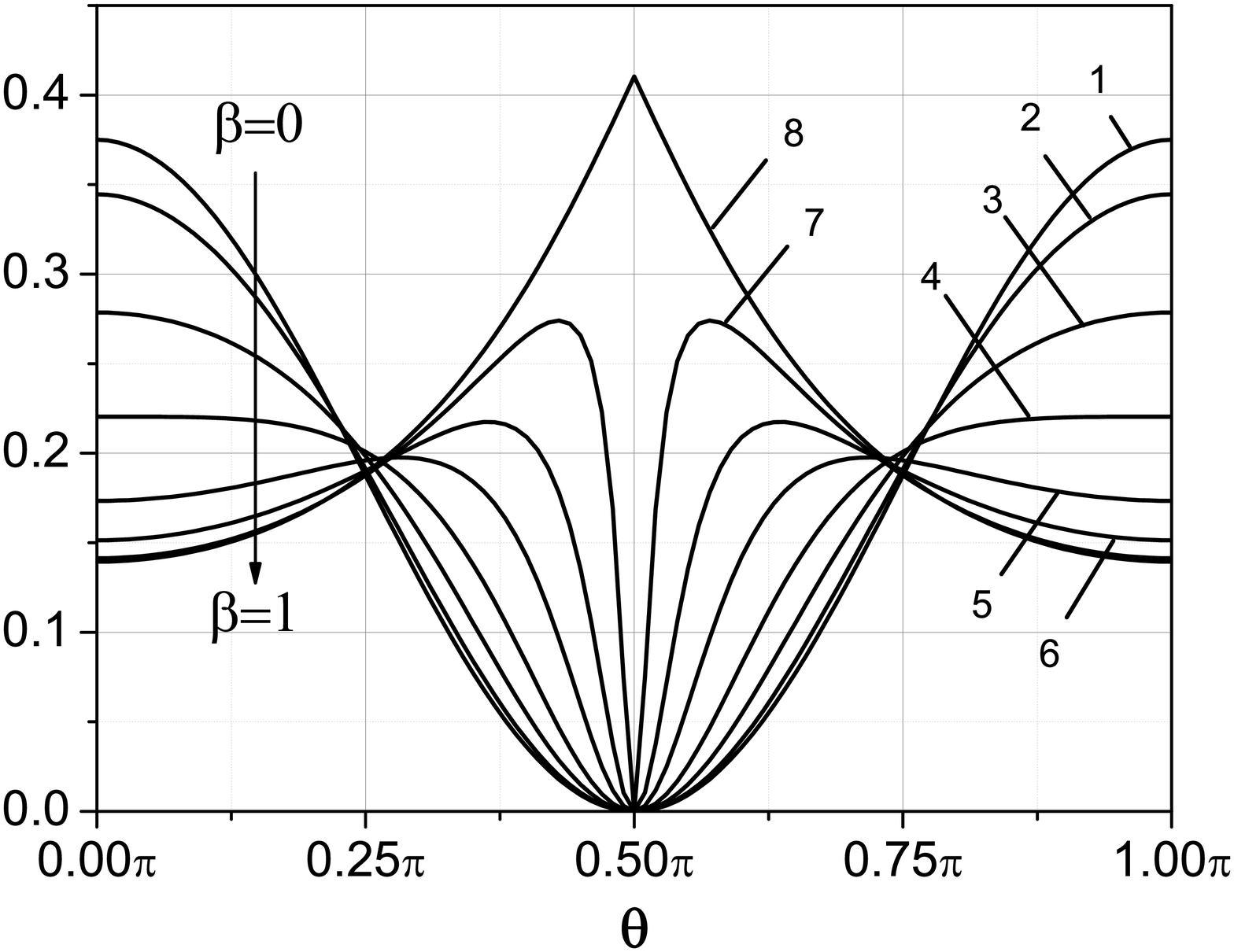}
\caption{Functions $p_{3}^{e}(\beta ;\theta )$. Curve numbers correspond to increasing values
of $\beta$ in the following order:\\ $\{ 0.0, 0.4, 0.7, \sqrt 3/2, 0.96, 0.99, 0.999, 1\}$.}
\end{minipage}
\end{figure*}

The dependence of functions $p_{2}^{b}(\beta ;\theta )$ and $p_{2}^{e}(\beta ;\theta )$ on $\beta $ has the simplest form (see the Fig.2 and 3). Moreover, both these functions are monotonically increasing functions of $\theta
$ in the region of $0\leqslant \theta \leqslant \frac{\pi }{2}$
(qualitatively this behavior is consistent with classical theory). Function $p_{2}^{b}(\beta ;\theta )$ is a monotonically decreasing function
of $\beta $ for each fixed $\theta $, whereas in classical theory there is
no such monotony. For function $p_{2}^{e}(\beta ;\theta )$ there is no
monotonous behavior on $\beta $, and though this shows the influence of particle spin, the behavior of the electronic functions is closer to classical theory than that of the bosonic functions.

The evolution of functions $p_{3}^{b}(\beta ;\theta )$ and $%
p_{3}^{e}(\beta ;\theta )$ is represented in Figures 4 and 5 respectively.
Function $p_{3}^{b}(\beta ;\theta )$ is monotonically decreasing with $%
\theta $ in the $0\leqslant \theta \leqslant \frac{\pi }{2}$ interval for
each fixed $\beta $, whereas function $p_{3}^{e}(\beta ;\theta )$ is
monotonically decreasing only for $\beta ^{2}\leqslant 3/4,\,\,(\gamma
\leqslant 2)$. For $\beta ^{2}>3/4,\,\,(\gamma >2)$, the functions are not monotonous any more, and an internal
maximum point appears for these functions at $\theta _{3}^{(max)}(\beta )$. For this case, the behavior of the electron distribution
functions is closer to classical theory compared to the boson ones.

Figures 6 and 7 show the evolution of the functions $p_{1}^{b}(\beta
;\theta )$ and $p_{1}^{e}(\beta ;\theta )$  in the $0\leqslant
\theta \leqslant \pi $ interval. From Fig. 6 it is clear that $%
p_{1}^{b}(\beta ;\theta )$ are monotonically decreasing functions of $\theta
$, whereas the functions $p_{1}^{e}(\beta ;\theta )$ for $\beta
^{2}>1/2,\,\,(\gamma ^{2}>2)$ lose their monotony and an internal maximum
point $\theta _{1}^{(max)}(\beta )$ appears for these functions, which also
qualitatively corresponds to classical theory. For $\beta \rightarrow 1$
the electron function approaches 0 in the $\frac{\pi }{2}\leqslant \theta
\leqslant \pi $ interval, which corresponds to the disappearance of the
right-handed polarization in the lower half plane.
\begin{figure*}[!]
\begin{minipage}[!]{8.1cm}
\centering
\includegraphics[width=8cm]{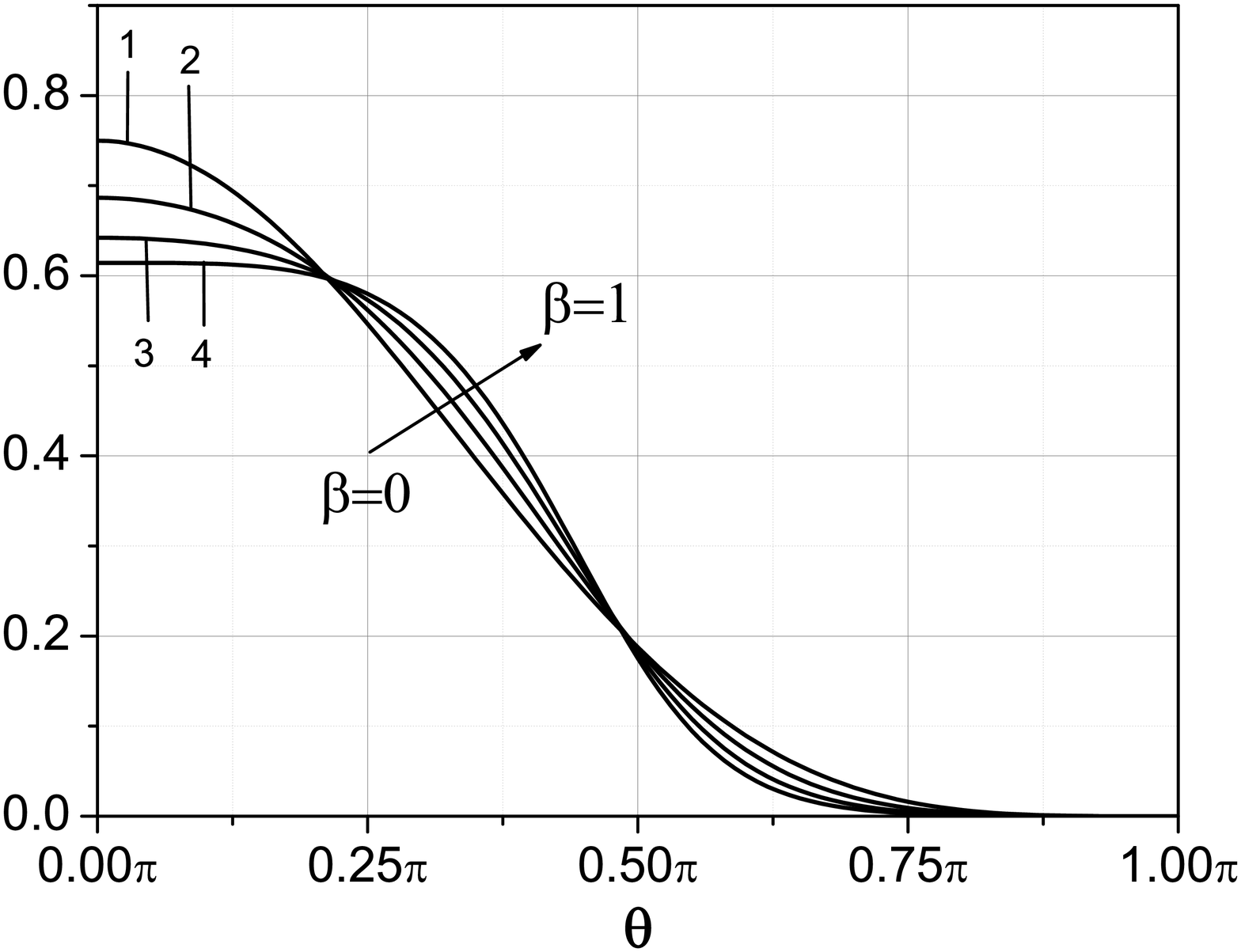}
\caption{ Functions $p_{1}^{b}(\beta ;\theta )$. Curve numbers correspond to increasing values
of $\beta$ in the following order:\\ $\{0.0, 0.7, 0.9, 1.0\}$.}
\end{minipage}
\hfill
\begin{minipage}[!]{8.1cm}
\centering
\includegraphics[width=8cm]{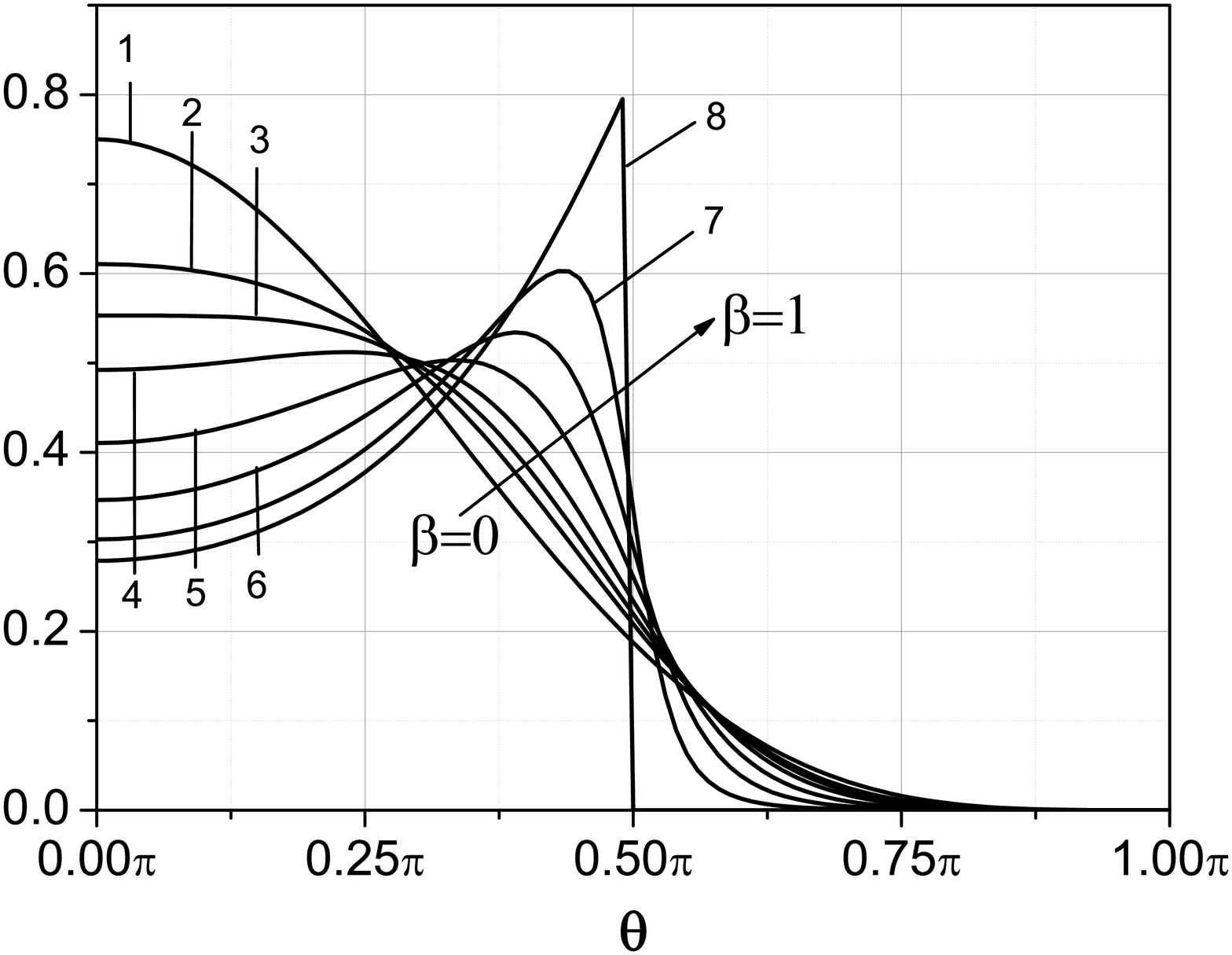}
\caption{Functions $p_{1}^{e}(\beta ;\theta )$. Curve numbers correspond to increasing values
of $\beta$ in the following order:\\ $\{ 0.0, 0.6, 0.1, 1/\sqrt 2, 0.8, 0.9, 0.96, 0.99, 0.999\}$.}
\end{minipage}
\end{figure*}

Lastly, the distribution functions of total radiation $
p_{0}^{b}(\beta ;\theta )$ and $p_{0}^{e}(\beta ;\theta )$ are shown on Fig.
8 and 9, from which it follows that the bosonic function is a
monotonically decreasing function of $\theta $ in the $0\leqslant \theta
\leqslant \frac{\pi }{2}$ interval (qualitatively the behavior is contrary
to classical theory). The behavior of electron functions is qualitatively
closer to classical theory and for small $\beta $ these functions are
monotonically decreasing. Still for $\beta ^{2}>1/2,\,\,(\gamma ^{2}>2)$
they lose monotony and there is an internal point of maximum for these functions at $\theta_{0}^{(max)}(\beta )$ .
\begin{figure*}[!]
\begin{minipage}[!]{8.1cm}
\centering
\includegraphics[width=8cm]{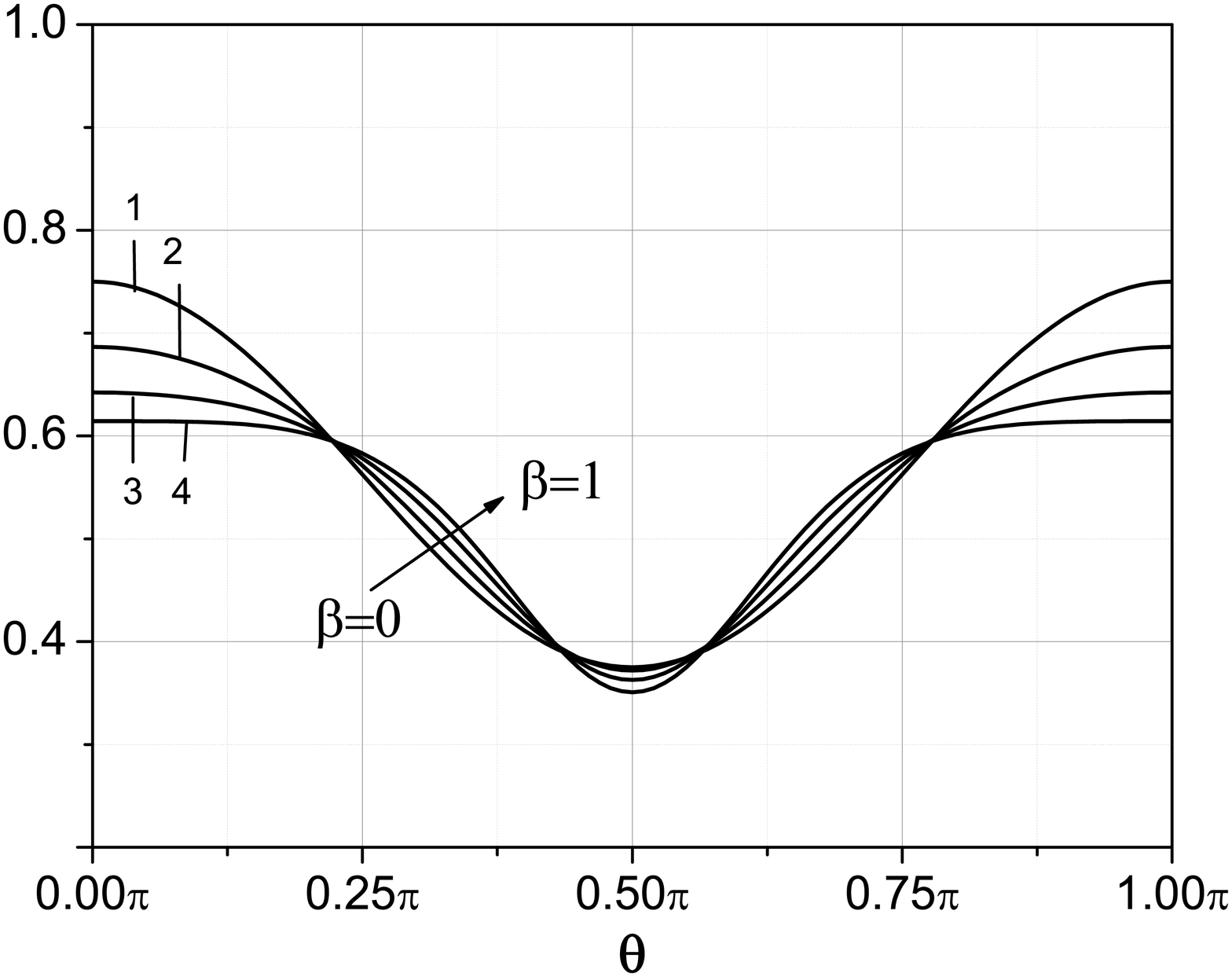}
\caption{Functions $p_{0}^{b}(\beta ;\theta )$. Curve numbers correspond to increasing values
of $\beta$ in the following order:\\ $\{0.0, 0.7, 0.9, 1\}$.}
\end{minipage}
\hfill
\begin{minipage}[!]{8.1cm}
\centering
\includegraphics[width=8cm]{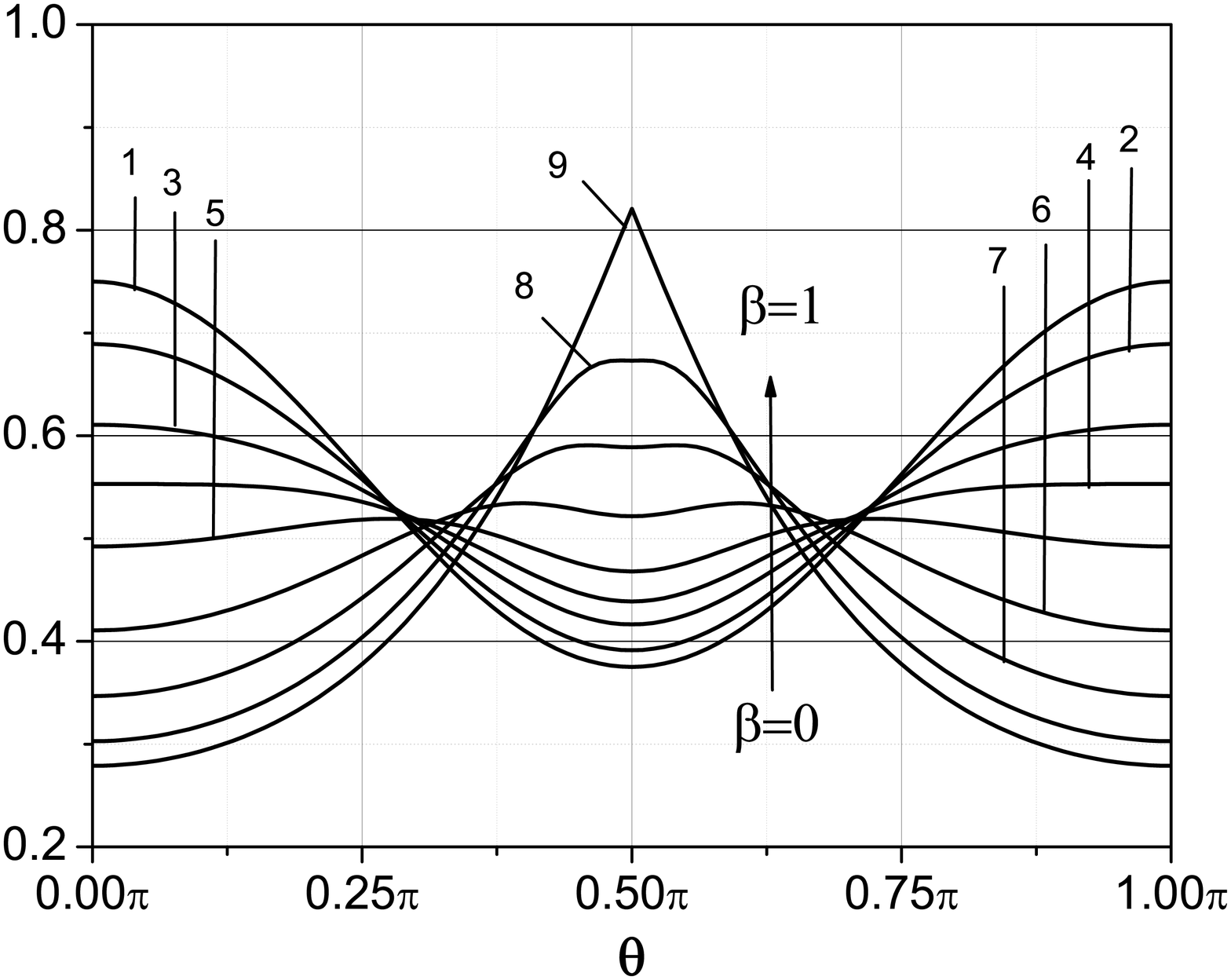}
\caption{Functions $p_{0}^{e}(\beta ;\theta )$. Curve numbers correspond to increasing  values
of $\beta$ in the following order:\\ $\{ 0.0, 0.4, 0.6, 1/\sqrt2, 0.8, 0.9, 0.96, 0.99, 1\}$.}
\end{minipage}
\end{figure*}

Figure 10 presents graphs of the functions $\theta _{s}^{(max)}(\beta
)\,\,(s=0,\,1,\,3)$. For $\beta \rightarrow 1\,\,(\gamma \rightarrow \infty
) $ all the angles $\ \theta _{s}^{(max)}(\beta )\rightarrow \frac{\pi }{2}$
depend on $\beta$ in the following way
\begin{eqnarray}
\theta _{0}^{max}(\beta )&\approx& \frac{\pi }{2}-\frac{2}{\gamma ^{2}},\quad
\theta _{1}^{max}(\beta )\approx \frac{\pi }{2}-\left( 2\gamma ^{2}\right)
^{-1/3},\quad
\theta _{3}^{max}(\beta )\approx \frac{\pi }{2}-\frac{1}{\sqrt{%
\gamma }}\,.\ \  \label{37}
\end{eqnarray}%
Figure 11 shows graphs of the maximum value for $p_{s}^{b(max)}(\beta
),\,\,p_{s}^{e(max)}(\beta )$.

Classical theory predicts the phenomenon of SR concentration at a narrow angle in the vicinity of the orbital plane for an ultra-relativistic particle. A necessary (but not sufficient) condition for this
concentration is the tendency of angles $\theta _{s}^{(max)}(\beta )$ to approach $\frac{\pi }{2}$ for $%
\beta \rightarrow 1\,\,(\gamma \rightarrow \infty )$. Although for an electron (and $%
\sigma $--components of the boson radiation) this condition is satisfied,
it is clear from the behavior of the functions $p_{s}^{b}(\beta ;\theta )$ and $%
p_{s}^{e}(\beta ;\theta )$, shown in the Fig.2 -- 9, that there
is not a significant concentration of radiated power in the orbital plane (in
the vicinity of $\theta =\frac{\pi }{2}$) in this case, which contradicts
classical theory.
\begin{figure*}[!]
\begin{minipage}[!]{8.1cm}
\centering
\includegraphics[width=8cm]{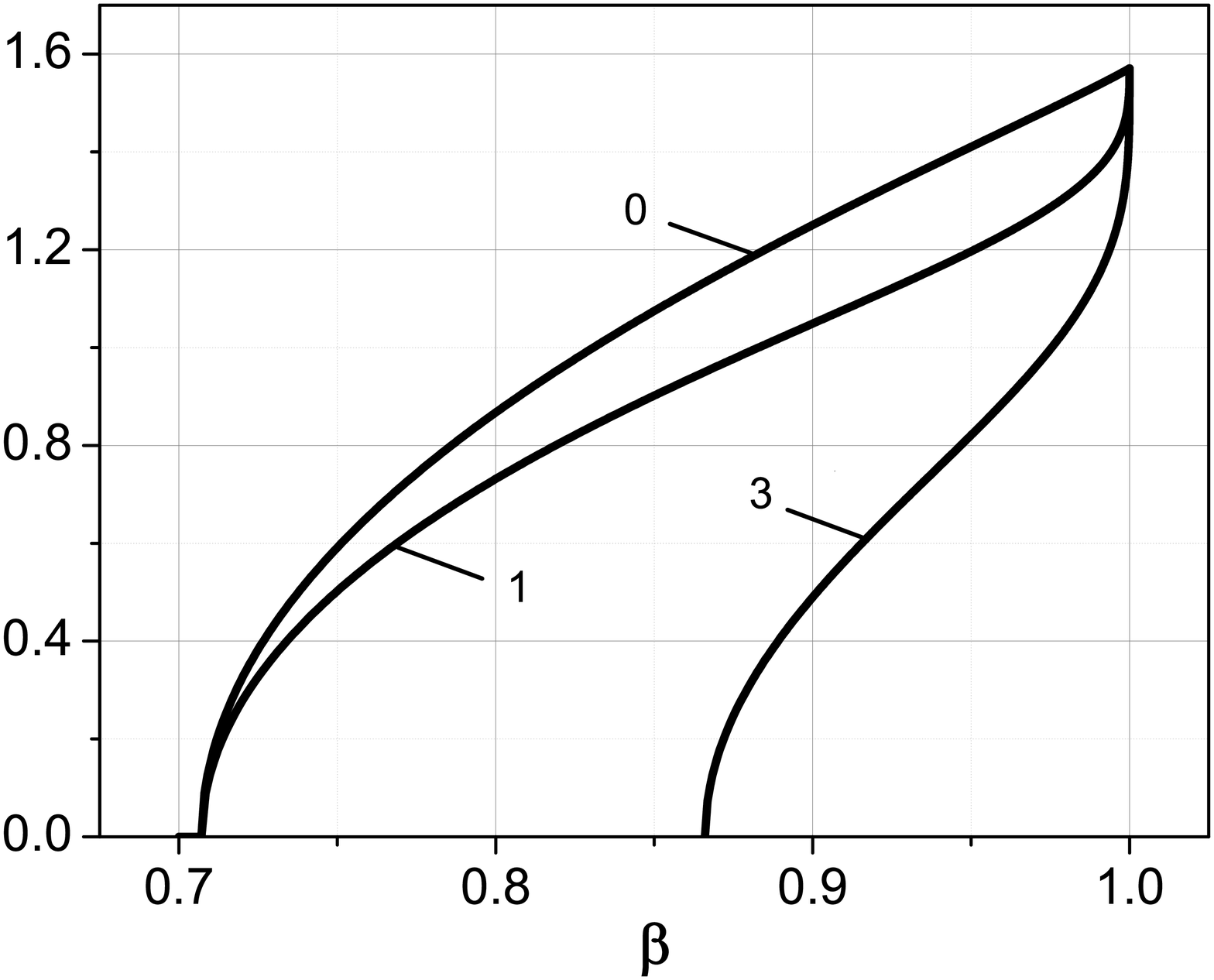}
\caption{Functions $\theta_{s}^{max}(\beta), (s=0, 1, 3)$.}
\end{minipage}
\hfill
\begin{minipage}[!]{8.1cm}
\centering
\includegraphics[width=8cm]{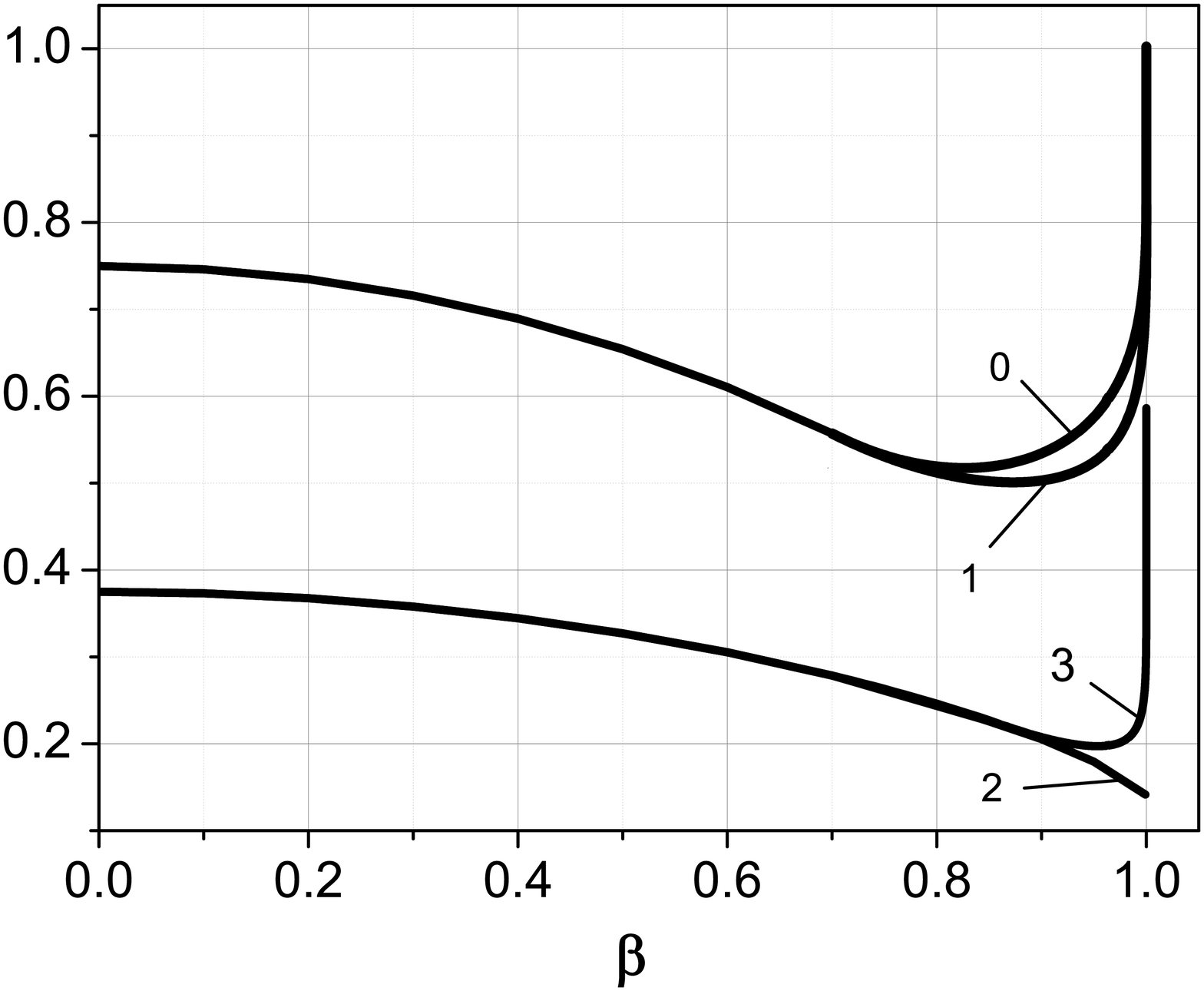}
\caption{Functions $p_{s}^{e(max)}(\beta), (s=0, 1, 2, 3)$.}
\end{minipage}
\end{figure*}

\begin{figure*}[!]
\begin{minipage}[!]{8.1cm}
\centering
\includegraphics[width=8cm]{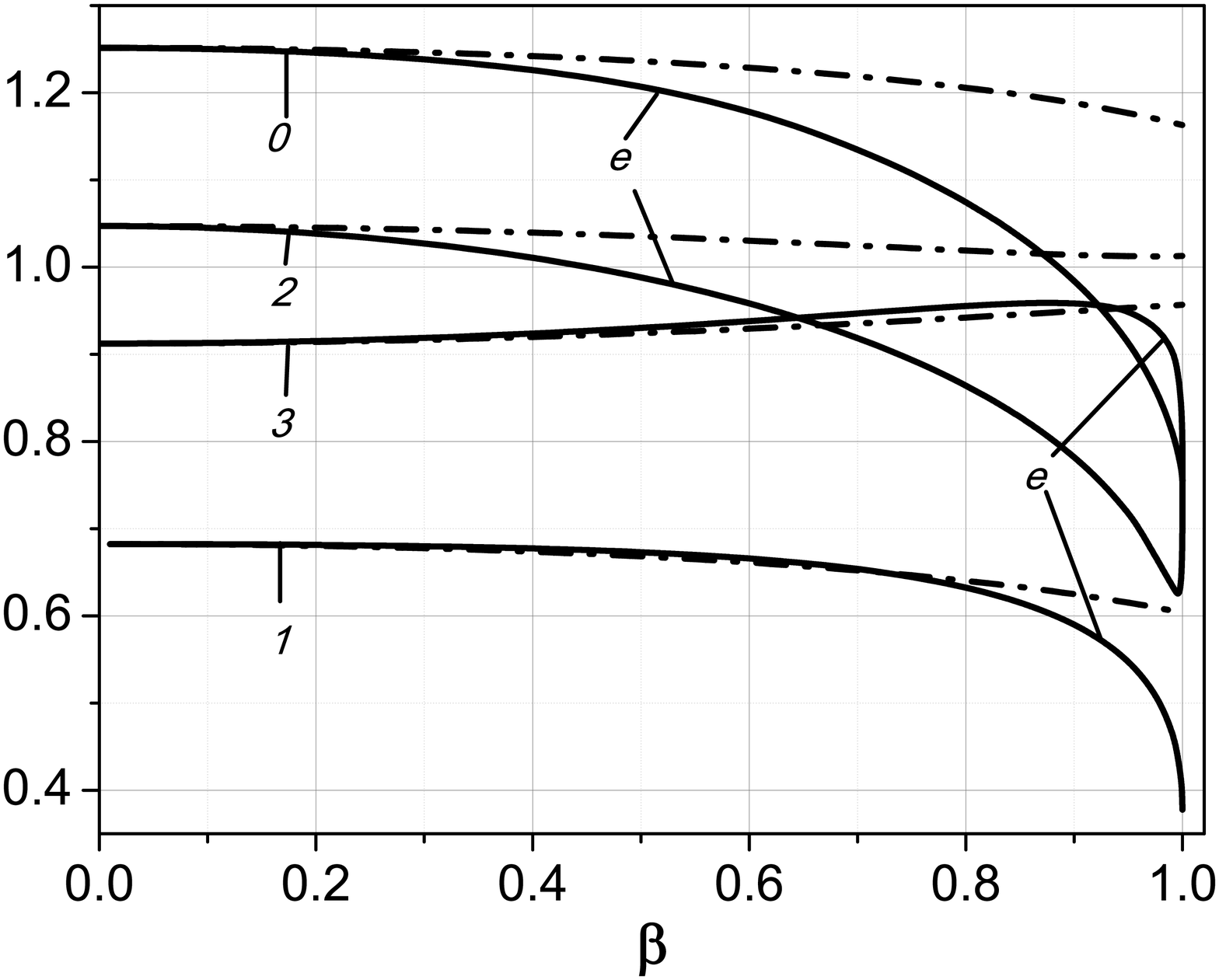}
\end{minipage}
\hfill
\begin{minipage}[!]{8.1cm}
\centering
\includegraphics[width=8cm]{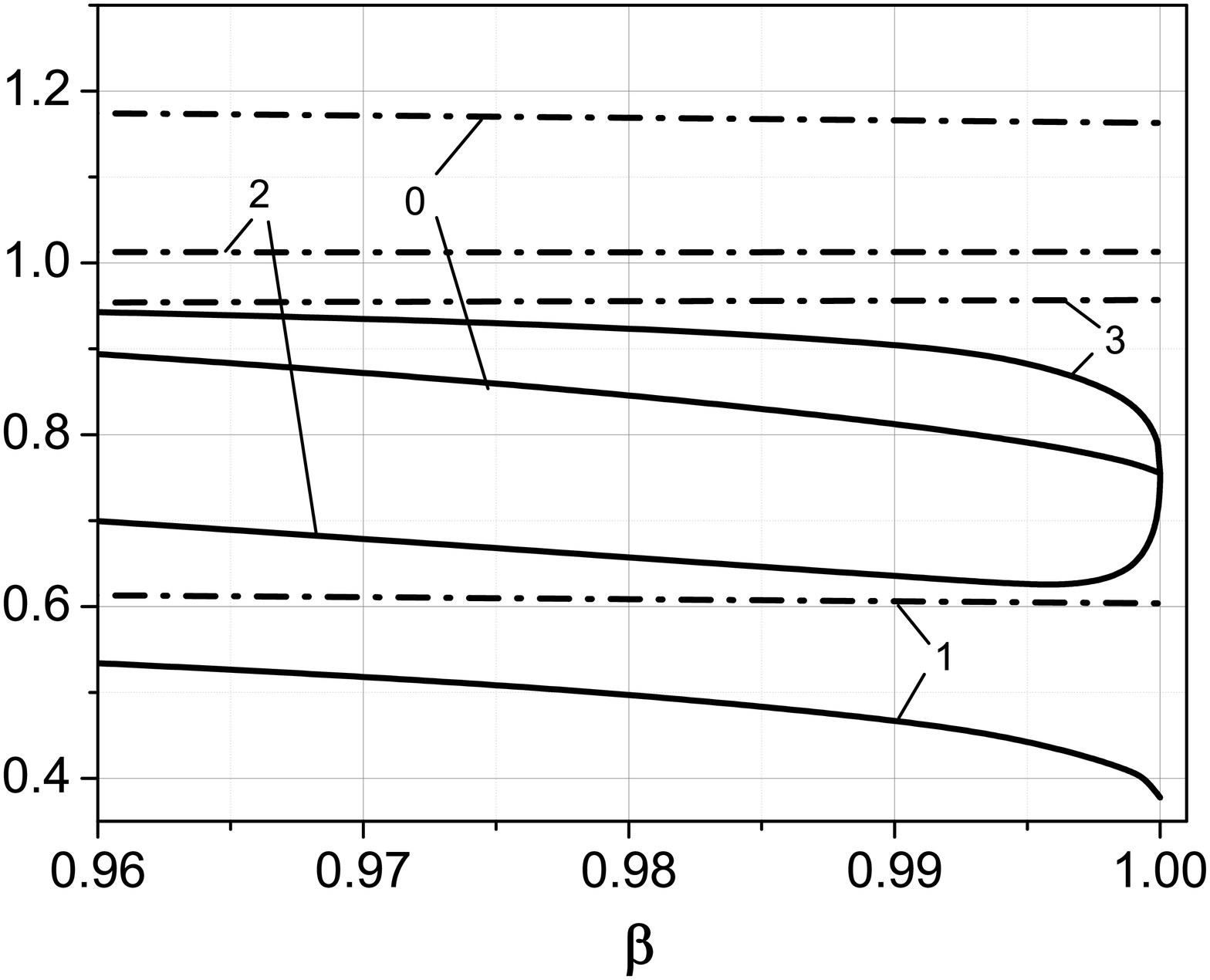}
\end{minipage}
\caption{Energy dependence of the for the effective angles $\Delta_{s}^{b}(\beta ),\,\,\Delta _{s}^{e}(\beta )$,$(s=0, 1, 2, 3)$.}
\end{figure*}

\begin{figure*}[!]
\begin{minipage}[!]{8.1cm}
\centering
\includegraphics[width=8cm]{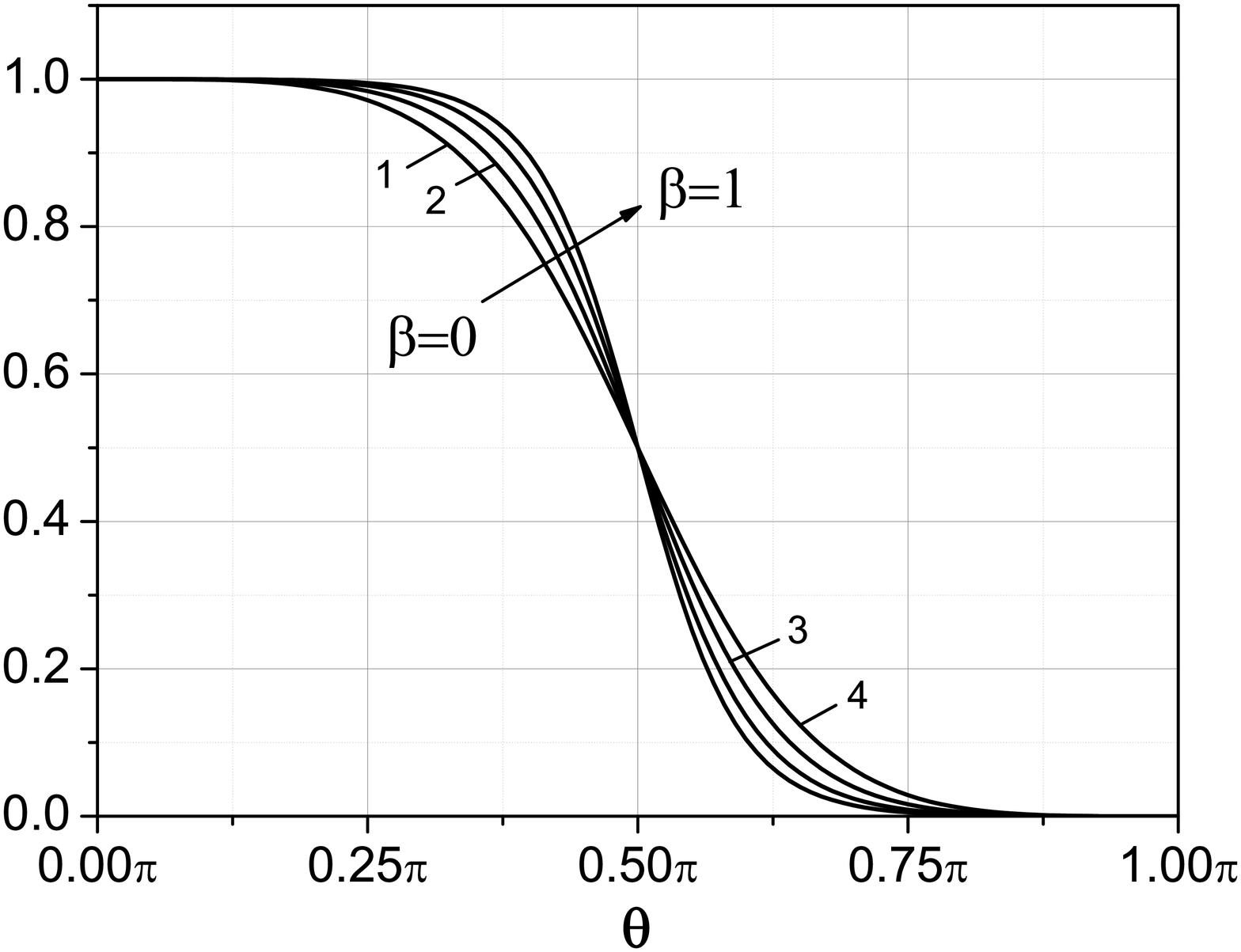}
\caption{Functions $q_{1}^{b}(\beta
;\,\theta )$. Curve numbers correspond to increasing values
of $\beta$ in the following order:\\ $\{0.0, 0.7, 0.9, 1.0\}$.}
\end{minipage}
\hfill
\begin{minipage}[!]{8.1cm}
\centering
\includegraphics[width=8cm]{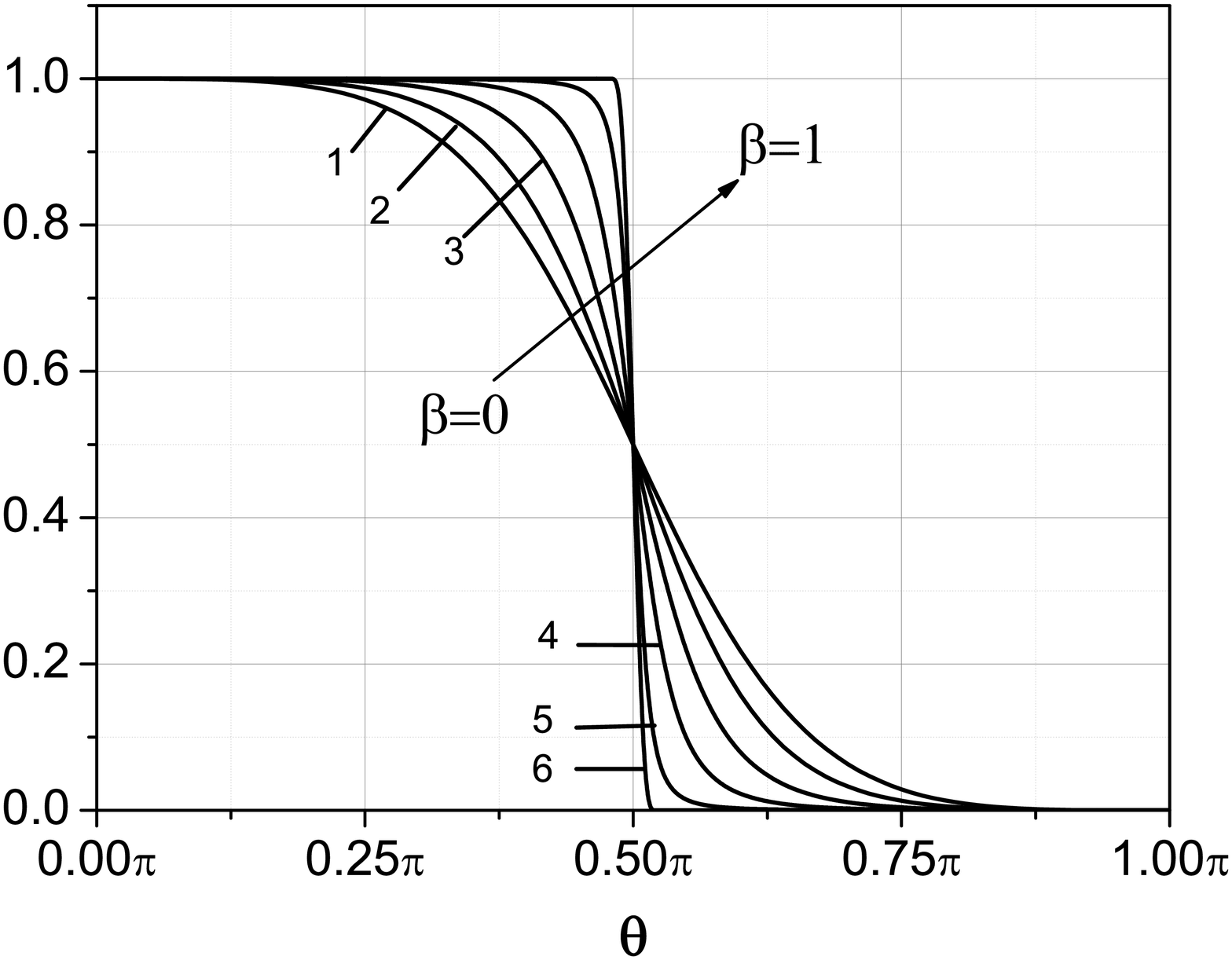}
\caption{Functions $q_{1}^{e}(\beta;\,\theta )$. Curve numbers correspond to increasing  values
of $\beta$ in the following order:\\ $\{ 0.0, 0.8, 0.95, 0.99, 0.999, 0.999999\}$.}
\end{minipage}
\end{figure*}

\begin{figure*}[!]
\begin{minipage}[!]{8.1cm}
\centering
\includegraphics[width=8cm]{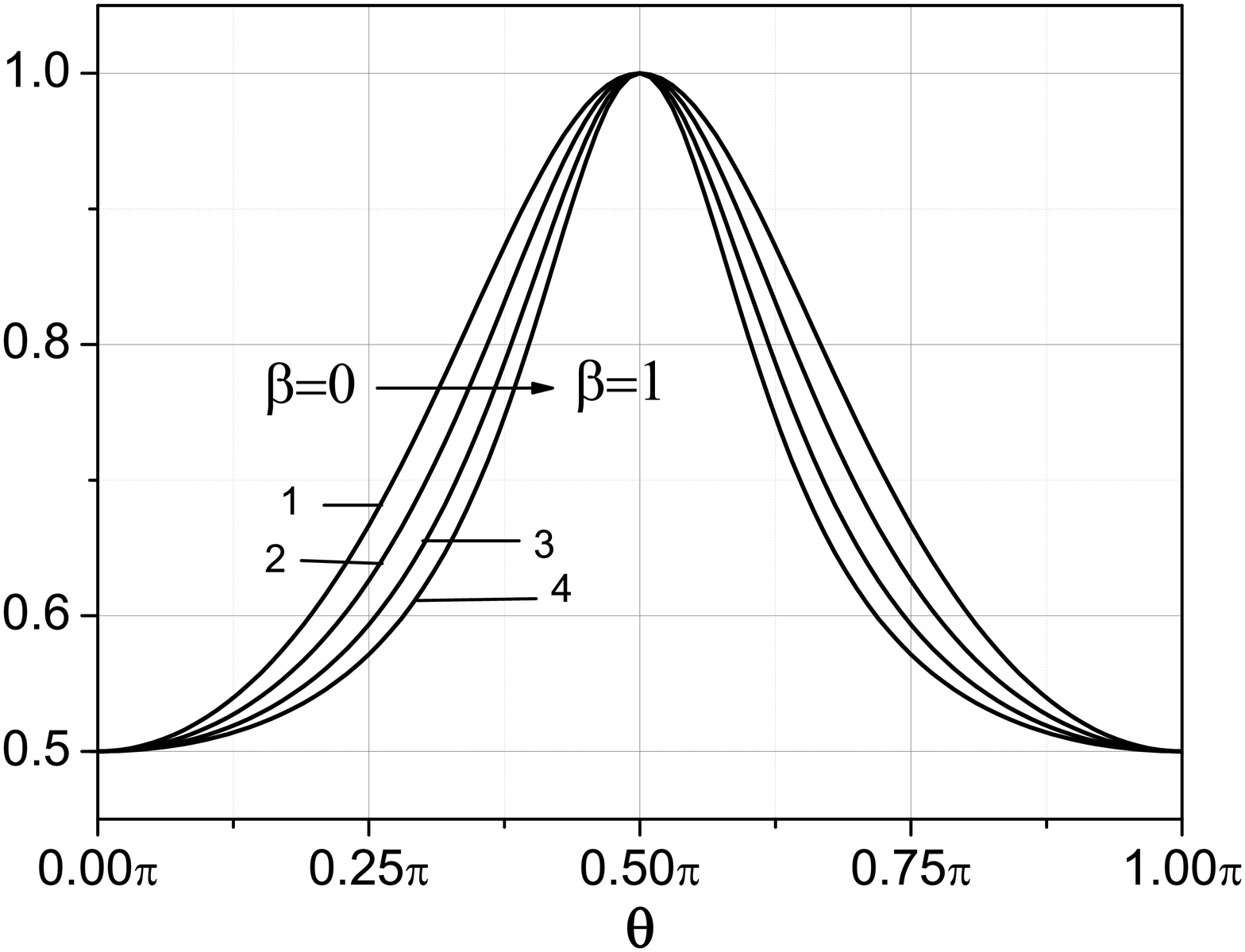}
\caption{Functions $q_{2}^{b}(\beta;\,\theta )$. Curve numbers correspond to increasing values
of $\beta$ in the following order:\\ $\{0.0, 0.7, 0.9, 1.0\}$.}
\end{minipage}
\hfill
\begin{minipage}[!]{8.1cm}
\centering
\includegraphics[width=8cm]{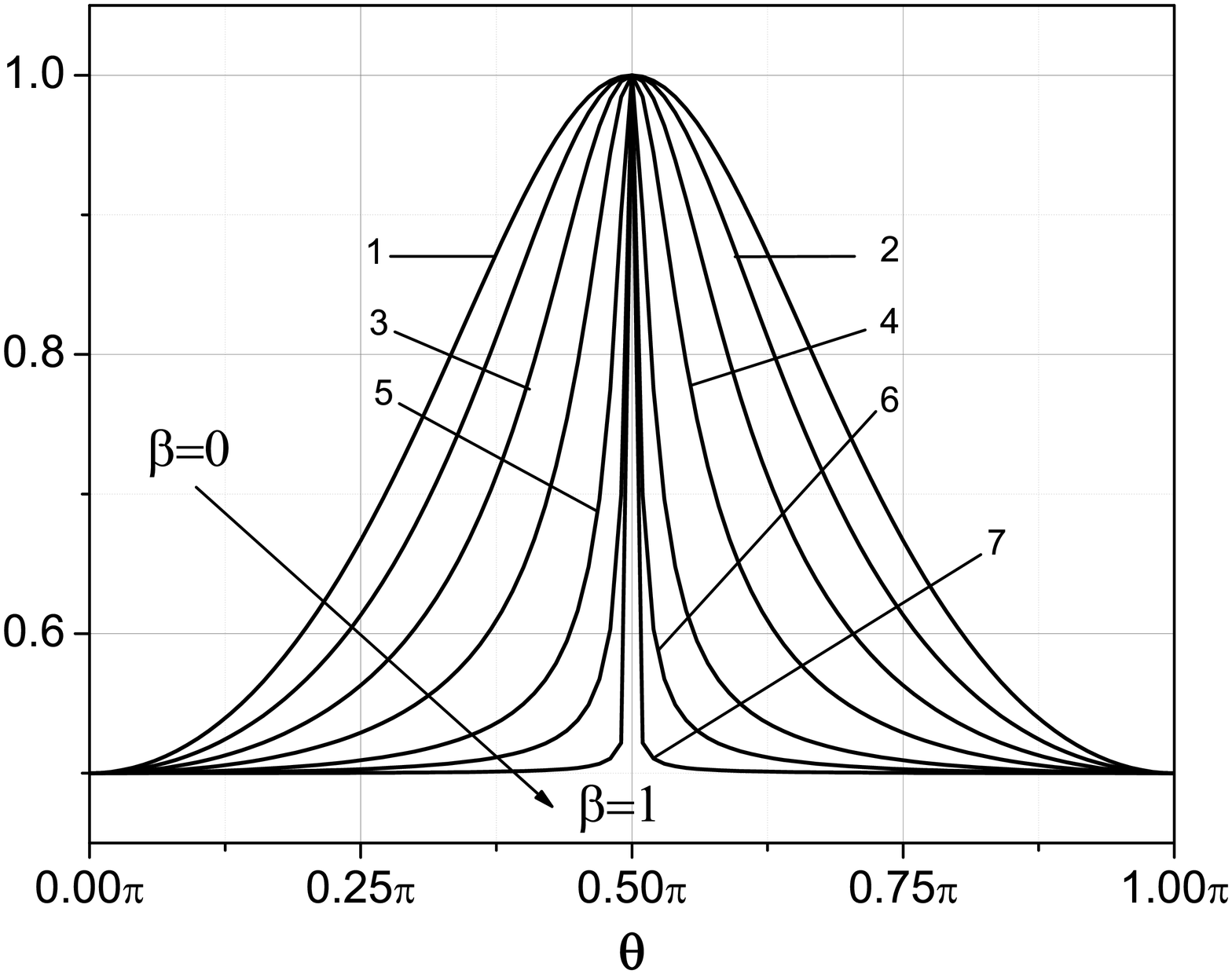}
\caption{Functions $q_{2}^{e}(\beta;\,\theta )$. Curve numbers correspond to increasing values
of $\beta$ in the following order:\\ $\{ 0.0, 0.8, 0.95, 0.99, 0.999, 0.9999, 0.999999\}$. }
\end{minipage}
\end{figure*}
\newpage

\textbf{7}. Lack of emission concentration in the vicinity of $\theta =\frac{\pi }{2}$ is also confirmed by the dependence of the effective angles $\Delta_{s}^{b}(\beta ),\,\,\Delta _{s}^{e}(\beta )$ on energy (the notion of an effective angle for SR was introduced in [17 -- 20]). Figure 12 shows graphs of these functions, from which it follows that for $s=0,\,1,\,2,\,$ these functions are weakly decreasing, tending to
a finite value at $\beta \rightarrow 1$. In the relativistic case, an arbitrarily large concentration of radiation in the orbit plane does not occur, in contrast to the conclusions of classical theory. The functions $%
\Delta _{3}^{b}(\beta ),\,\,\Delta _{3}^{e}(\beta )$ are increasing with
$\beta $, which means that the $\pi$--component of emission is diverging.

\textbf{8}. The most qualitative agreement with the results of classical theory is observed in the evolution of functions $q_{s}^{b}(\beta ;\,\theta ),\,\,q_{s}^{e}(\beta ;\,\theta ),$ $(s=1,\,2)$, which determine
the polarization of radiation for each fixed $\beta $ at an angle $\theta $.
Graphs of these functions are shown in Figures 13 -- 16. From these graphs it follows that in the field direction ($\theta =0$), the radiation has completely right-handed polarization and preferential linear polarization in this direction is missing. In
the orbital plane ($\theta =\frac{\pi }{2}$) there is a complete linear
polarization, and the preferential circular one is absent, which coincides with
the conclusions of classical theory. Quantitative differences between $q_{s}^{b}(\beta ;\,\theta )$ and $q_{s}^{e}(\beta ;\,\theta ),\,\,(s=1,\,2)$ become noticeable only for $\beta \rightarrow 1$.

\section*{\protect Acknowledgments}

The work of VGB was partially supported by the Federal Targeted
Program "Scientific and scientific - pedagogical personnel of innovative
Russia", contract No  P789; DMG acknowledges the permanent
support of FAPESP and CNPq; ANB acknowledges the permanent
support of FAPESP.

\end{document}